\documentclass{article}

\usepackage[margin=1in]{geometry}
\usepackage{charter}
\usepackage{amsmath,amssymb}
\usepackage{amsthm}
\usepackage{color,xcolor,graphicx}
\usepackage{algorithm,algorithmicx}
\usepackage{thm-restate}
\usepackage{boxedminipage, wrapfig}

\usepackage[pdftex,pagebackref,colorlinks,linkcolor=blue,filecolor = blue, citecolor = blue, urlcolor  = blue]{hyperref}

\newcommand{\expn}[3]{\phi^{#1}_{#2}\paren{#3}}
\newcommand{\expnG}[2]{\phi^{#1}_{#2}}
\newcommand{\uexpn}[2]{\phi^{\leftrightarrow}_{#1}\paren{#2}}
\newcommand{\uexpnG}[1]{\phi^{\leftrightarrow}_{#1}}
\newcommand{\spar}{\Phi}
\newcommand{\allones}{\mathbf{1}}
\newcommand{\vol}[2]{{\sf vol}_{#1}\paren{#2}}
\newcommand{\wt}[2]{{\sf wt}_{#1}\paren{#2}}

\newcommand{\R}{\mathbb{R}}
\newcommand{\Rplus}{\R_{+}}
\newcommand{\Rgeq}{\R_{\geq 0}}
\newcommand{\cR}{\mathcal{R}}
\newcommand{\cN}{\mathcal{N}}
\newcommand{\defeq}{\overset{\text{\tiny def}}{=}}
\newcommand{\Abs}[1]{\left|#1\right|}
\newcommand{\set}[1]{\left\{#1\right\}}
\newcommand{\paren}[1]{\left(#1\right)}
\newcommand{\sqbracs}[1]{\left[#1\right]}
\newcommand{\bigo}[1]{O\paren{#1}}
\newcommand{\inG}[1]{{#1}^{(G)}}

\newcommand{\Ex}[2]{\underset{#1}{\mathbb{E}}\left[#2\right]}
\newcommand{\Prob}[2]{\underset{#1}{\mathbb{P}}\left[#2\right]}
\newcommand{\inprod}[1]{\left\langle#1\right\rangle}
\newcommand{\norm}[1]{\left\|#1\right\|}
\newcommand{\e}{\epsilon}
\newcommand{\fplus}[2]{f^+_{#1}\paren{#2}}
\newcommand{\fminus}[2]{f^-_{#1}\paren{#2}}
\newcommand{\rhoplus}[2]{\rho^+_{#1}\paren{#2}}
\newcommand{\rhoplussymb}[1]{\rho^+_{#1}}
\newcommand{\rhominus}[2]{\rho^-_{#1}\paren{#2}}
\newcommand{\rhominussymb}[1]{\rho^-_{#1}}
\newcommand{\hrhosymb}{\hat{\rho}}
\newcommand{\hrho}[2]{\hat{\rho}_{#1}\paren{#2}}
\newcommand{\hrhoplus}[2]{\hat{\rho}^+_{#1}\paren{#2}}
\newcommand{\hrhominus}[2]{\hat{\rho}^-_{#1}\paren{#2}}
\newcommand{\cutvalsymb}{{\sf cutval}}
\newcommand{\cut}[2]{{\sf cut}_{#1}\paren{#2}}
\newcommand{\cutval}[2]{{\sf cutval}_{#1}\paren{#2}}
\newcommand{\symcut}[2]{\overleftrightarrow{{\sf cut}}_{#1}\paren{#2}}
\newcommand{\symcutval}[2]{\overleftrightarrow{{\sf cutval}}_{#1}\paren{#2}}
\newcommand{\vect}[1]{{\sf v}_{#1}}
\newcommand{\dv}[4]{{\bf d}_{#2,#3,#4}^{#1}}
\newcommand{\dvright}[4]{{\bf d}_{#2,#3,#4}^{R,#1}}
\newcommand{\dvleft}[4]{{\bf d}_{#2,#3,#4}^{L,#1}}
\newcommand{\supp}[1]{{\sf supp}\paren{#1}}
\newcommand{\lovasz}{Lov\'asz}
\newcommand{\sdpval}{{\sf SDP}}
\newcommand{\opt}{{\sf OPT}}
\newcommand{\ddistsymb}{\overrightarrow{\ell}}
\newcommand{\ddist}[1]{\ddistsymb\paren{#1}}
\newcommand{\vectone}{\vect{\emptyset}}

\newcommand{\dr}{dr}
\newcommand{\dt}{dt}
\newcommand{\Ind}[1]{\mathbb{I}\paren{#1}}
\newcommand{\argmin}{{\sf argmin}}
\newcommand{\ral}[3]{\cR_{#1}^{#2}\paren{#3}}
\newcommand{\ralu}[2]{\cR_{#1}^{\leftrightarrow}\paren{#2}}
\newcommand{\nin}[2]{N_{#1}^{{\sf in}}\paren{#2}}
\newcommand{\nout}[2]{N_{#1}^{{\sf out}}\paren{#2}}

\newcommand{\basepoly}[1]{\mathcal{B}_{#1}}
\newcommand{\cP}{\mathcal{P}}
\newcommand{\cT}{\mathcal{T}}

\theoremstyle{plain}
\newtheorem{theorem}{Theorem}[section]

\theoremstyle{definition}
\newtheorem{proposition}[theorem]{Proposition}
\newtheorem{sdp}[theorem]{SDP}
\newtheorem{lemma}[theorem]{Lemma}
\newtheorem{claim}[theorem]{Claim}
\newtheorem{corollary}[theorem]{Corollary}
\newtheorem{definition}[theorem]{Definition}
\newtheorem{fact}[theorem]{Fact}

\newtheorem{remark}[theorem]{Remark}

\title{On Sparsest Cut and Conductance \\ in Directed Polymatroidal Networks}

\author{
Chandra Chekuri\thanks{Work on this was made possible by visits to the Indian Institute of Science, Bengaluru
supported by a Smt.\ Rukmini and Shri.\
Gopalakrishnachar Visiting Professorship. Work at Univ.\ of Illinois supported in part by NSF grant CCF-2402667.} \\
University of Illinois, \\
Urbana-Champaign \\
chekuri@illinois.edu
\and 
Anand Louis\thanks{Supported in part by SERB Award CRG/2023/002896 and the Walmart Center for Tech Excellence at IISc (CSR Grant WMGT23-0001).} \\
Indian Institute of Science, \\
Bengaluru \\
anandl@iisc.ac.in
}

\date{}
\begin{document}
\begin{titlepage}

\maketitle

\begin{abstract}
  We consider algorithms and spectral bounds for sparsest cut and
  conductance in directed polymatrodal networks. This is motivated
  by recent work on submodular hypergraphs
  \cite{Yoshida19,LiM18,ChenOT23,Veldt23} 
  and previous work on multicommodity flows and cuts in polymatrodial
  networks \cite{ChekuriKRV15}. We obtain three results.
  First, we obtain an $O(\sqrt{\log n})$-approximation for sparsest
  cut and point out how this generalizes the result in
  \cite{ChenOT23}. Second, we consider the symmetric version of
  conductance and obtain an $O(\sqrt{\opt \log r})$ approximation
  where $r$ is the maximum degree and we point out how this
  generalizes previous work on vertex expansion in graphs.
  Third, we prove a non-constructive Cheeger like inequality that
  generalizes previous work on hypergraphs. We provide a unified
  treatment via line-embeddings which were shown to be effective
  for submodular cuts in \cite{ChekuriKRV15}.
\end{abstract}

\end{titlepage}

\tableofcontents

\section{Introduction}
\label{sec:intro}
Flows, cuts, and partitioning problems in graphs have a number of
applications and have been extensively studied in the literature. We
will be primarily interested in the study of expansion and sparsest
cut problems.  These have several applications and connections to
various topics in algorithms, optimization and mathematics.  Recently
there has been interest in the study of these problems in
\emph{submodular} hypergraphs
\cite{Yoshida19,LiM18,Veldt23,ChenOT23}. Some of this work has
connections to past work on multicommodity flows and cuts in
polymatrodial-networks \cite{ChekuriKRV15}. The work in
\cite{ChekuriKRV15} was mostly focused on flow-cut gaps and related
algorithmic results --- the impetus for that work came form
information theory considerations \cite{AvestimehrDS11,EbrahimiF12,YazdiS2011,GoemansIZ12,KannanV14,RajaV11}. In the conclusions of
\cite{ChekuriKRV15} it was suggested that SDP based approximation
algorithms may also be feasible for sparsest cut and related problems
in such networks. Recent work on submodular hypergraphs has relied on
SDPs and spectral methods (among others).  In particular
\cite{ChenOT23} obtained an $O(\sqrt{\log n})$-approximation for the
sparsest cut problem in a certain class of submodular hypergraphs. In
this work we examine these techniques and show that the line-embedding
ideas from \cite{ChekuriKRV15} are useful in proving several results.

\paragraph{Submodular set functions:} A real-valued set function
$f:2^N \to \mathbb{R}$ over a finite ground set $N$ is submodular if
$f(A) + f(B) \ge f(A \cup B) + f(A \cap B)$ for all
$A, B \subseteq N$. Other equivalent definitions via diminishing
marginal values are known and we discuss them in Section
\ref{sec:background}.  We will assume that all submodular functions in
this paper are normalized, that is, they satisfy $f(\emptyset) =
0$. $f$ is monotone if $f(A) \le f(B)$ for all $A \subseteq B$. $f$ is
symmetric if $f(A) = f(N \setminus A)$ for all $A$. We restrict
attention to non-negative submodular functions.

\paragraph{Polymatroidal Networks:} Independent work of Hassin
\cite{Hassin78} and Lawler and Martel \cite{LawlerM82} defined the
notion of polymatrodial networks which is closely related to the work
of Edmonds and Giles on submodular flows \cite{EdmondsG77}. These
classical works generalized the standard notion of $s$-$t$ flows and
cuts and circulations, and have a number of applications in
combinatorial optimization.  Chekuri et al.\ \cite{ChekuriKRV15}
studied the generalization of polymatrodal networks to the
\emph{multicommodity} setting and we follow their notation and
definitions since we build on their work. A directed polymatroidal
network consists of a directed graph $G=(V,E)$ along with two monotone
submodular functions $\rhominussymb{v}$ and $\rhoplussymb{v}$
associated with each vertex $v \in V$.  Let $\delta^-(v)$ and
$\delta^+(v)$ denote the arcs into $v$ and out of $v$
respectively. Then $\rhominussymb{v}:2^{\delta^-(v)} \to \Rgeq$
and $\rhoplussymb{v}:2^{\delta^+(v)} \to \Rgeq$ are functions
that assign joint capacity constraints on the arcs coming into $v$ and
leaving $v$ respectively. In \cite{ChekuriKRV15} undirected
polymatrodial networks were defined. Here the graph $G=(V,E)$ is
undirected and there is a monotone submodular function
$\rho_v:2^{\delta(v)} \to \Rgeq$ associated with each $v \in V$
where $\delta(v)$ is the set of edges incident to $v$.  We assume that
the functions $\rhoplussymb{v},\rhominussymb{v} \ v \in V$, are
provided via value oracles.

\paragraph{Cut and Partitioning Problems}
A number of cut and partitioning problems are studied in graphs and
the can be extended naturally to hypergraphs and also to polymatroidal
networks.  A cut refers to a subset of edges. Sometimes we refer to a set of 
vertices $S \subseteq V$ as a cut in which case we mean $\delta(S)$.
The definition of the value of the cut, denoted by $\cutvalsymb$,
depends on the problem. For undirected graphs and
hypergraph $\cutval{}{S} = |\delta_G(S)|$. When edges have capacities $c:E \to \Rplus$, 
we consider $\cutval{}{\delta_G(S)} \defeq \sum_{e \in \delta_G(S)}
c(e)$.

The definition of the value of a cut in a polymatroidal networks is
more involved and is defined as follows. 
\[ \cutval{}{S} \defeq \min_{F \in \set{\delta^+(S), \delta^-(S)}} \min_{g \in \set{F \to V}} 
 \sum_{v \in V} \paren{\rhoplus{v}{\delta^+(v) \cap g^{-1}(v)} 
	+ \rhominus{v}{\delta^-(v) \cap g^{-1}(v)}}.\]
The ``symmetric'' value of the cut is defined as follows. 
\[ \symcutval{}{S} \defeq \sum_{F \in \set{\delta^+(S), \delta^-(S)}} \min_{g \in \set{F \to V}} 
 \sum_{v \in V} \paren{\rhoplus{v}{\delta^+(v) \cap g^{-1}(v)} 
   + \rhominus{v}{\delta^-(v) \cap g^{-1}(v)}}.\]
The symmetric formulation also enables us to
address undirected problems in a clean fashion.
We discuss this in more detail in Section \ref{sec:polymatroidal}.
In this paper we are mainly concerned with the sparsest
cut problem and the conductance problem.

\begin{definition}[Sparsest cut]
\label{def:sparsity}
Let $w: V \to \Rgeq$ be non-negative weights
on the vertices. 
The sparsity of a set $S \subseteq V$ is defined as
$\spar(S) \defeq \frac{\cutval{}{S}}{w(S)w(V \setminus S)}$. 
The sparsity of $G$ is defined as $\spar_G \defeq \min_{S \subseteq V} \spar(S)$.	
The sparsest cut is defined as $\argmin_{S \subseteq V} \spar(S)$.
\end{definition}

\paragraph{Conductance:} 
Conductance is defined with respect to ``volume'' of sets which is defined as
$\vol{}{S} \defeq \sum_{v \in S} \deg_v$. 
For graphs and hypergraphs, the degree of a vertex $v$ is defined as
the sum of weights of the hyperedges it belongs to. For polymatroidal 
networks, we define $\deg_v \defeq \rhoplus{v}{\delta^+(v)} 
+ \rhominus{v}{\delta^-(v)}$. 
\begin{definition}[Conductance]
\label{def:conductance}
The conductance of a set $S \subseteq V$, denoted by $\expn{}{G}{S}$, and \emph{symmetric} conductance,
denoted by $\uexpn{G}{S}$, are defined as
\[ \expn{}{G}{S} \defeq
\frac{\cutval{}{S}}{\min\set{\vol{}{S},\vol{}{V \setminus S}}} 
\qquad \textrm{and} \qquad
\uexpn{G}{S} \defeq
\frac{\symcutval{}{S}}{\min\set{\vol{}{S},\vol{}{V \setminus S}}} . \]
The conductance and the symmetric conductance of $G$, denoted by $\expnG{}{G}$
and $\uexpnG{G}$ respectively, are defined as 
\[ \expnG{}{G} \defeq \min_{S \subset V} \expn{}{G}{S} 
\qquad \textrm{and} \qquad
\uexpnG{G} \defeq \min_{S \subset V} \uexpn{G}{S} . \]
\end{definition}

One motivation for symmetric conductance comes from undirected graphs (both edge and
vertex capacitated) and hypergraphs. It is feasible to work with them separately but it is helpful to unify 
these via analysis on directed graphs.
We discuss more details in Section \ref{sec:polymatroidal}.

\paragraph{Submodular and Polymatroidal Hypergraphs:} A hypergraph $G=(V,E)$ consists of
a finite vertex set $V$ and a set of hyper-edges $E$ where each hyperedge $e \subseteq V$
is a subset of vertices. The rank of a hypergraph, denoted by $r$, is $\max_{e \in E} |e|$ and a hypergraph
is $k$-uniform if $|e| = k$ for all $e  \in E$. Graphs without self-loops are precisely $2$-uniform hypergraphs.
Cuts and partition problems on graphs can and have been generalized to hypergraphs; there are close
connections between hypergraphs and vertex-weighted problems in graphs via the bipartite/factor graph
representation of hypergraphs \cite{Lawler73}. Yoshida \cite{Yoshida19} and Li and Milenkovic \cite{LiM18}
independently considered submodular hypergraphs. In this model each hyperedge $e \in E$ is associated with
a submodular function $f_e:2^e \rightarrow \mathbb{R}_+$. One can define sparse cuts in this model
and the main contribution of \cite{Yoshida19,LiM18} are Cheeger-style inequalities based on spectral methods.
There are computational barriers to finding good approximations for these problems due to lower bounds stemming from \cite{SvitkinaF11}
and the results in \cite{Yoshida19,LiM18} provide weak approximation bounds for solving the underlying relaxations.
In more recent work, Chen, Orecchia and Tani \cite{ChenOT23} and Veldt \cite{Veldt23} have considered special cases of
submodular hypergraphs and derived provable approximation algorithms for sparsest cut via SDP and LP based methods and
also developed fast algorithms via the cut-matching approach \cite{KRV,OSSV,OrecchiaATT22}. In particular, \cite{ChenOT23} defined the
class of polymatroidal hypergraphs which are closely related to the polymatroidal network model. We give more details
in Section~\ref{sec:related}. One of the motivations for this paper is to generalize the SDP based and spectral based
methods to (directed) polymatrodal networks as a way to unify several results in the literature.

Our focus in this paper is on polynomial-time approximation algorithms and we do not address the development of faster approximation algorithms. We hope to investigate such algorithms in the future.

\subsection{Our results}
We will use $n$ to denote the number of nodes in a given graph.
We let $r$ denote the maximum degree of the graph (the degree of
a node is the sum of its in and out degrees).

Our first result is $\bigo{\sqrt{\log n}}$ approximation algorithm for 
$\spar_G$ of polymatroidal networks generalizing the corresponding result
for graphs \cite{AroraRV09}, vertex-weighted graphs \cite{FeigeHL08}, hypergraphs \cite{LouisM16}, directed
graphs \cite{AgarwalCMM05}, and polymatroidal  hypergraphs \cite{ChenOT23}.
We prove Theorem \ref{thm:mult-approx} in Section \ref{sec:mult-approx}.

\begin{restatable}{theorem}{multapprox}
\label{thm:mult-approx}
There is a polynomial time $\bigo{\sqrt{\log n}}$-approximation algorithm for 
$\spar_G$ of a directed polymatroidal network
$G = \paren{V,E,(w_i)_{i \in V},(\rhoplussymb{v},\rhominussymb{v})_{v
    \in V}}$. 
\end{restatable}

\begin{remark}
  The ratio can be shown to be $\bigo{\sqrt{\log k}}$ where $k$ is the cardinality of the support of $w$.
\end{remark}

We note that an $O(\log n)$-approximation via flow-cut gap is known
from \cite{ChekuriKRV15}.

Next, we give a $\bigo{\sqrt{\uexpnG{G} \log r}}$ bound for the symmetric
conductance of a polymatroidal network.
We prove Theorem \ref{thm:alg1} in Section \ref{sec:alg1}.
\begin{restatable}{theorem}{talgcheeger}
\label{thm:alg1}
Let $G = \paren{V,E,(w_i)_{i \in V},(\rhoplussymb{v},\rhominussymb{v})_{v \in V}}$
be a directed polymatroidal network, and let its maximum degree
be denoted by $r \defeq \max_{v \in V} \paren{\Abs{\delta^+(v)} +
  \Abs{\delta^-(v)}}$.
There is a randomized polynomial time algorithm that outputs a set $S \subset V$ 
such that $\uexpn{G}{S} = \bigo{\sqrt{\uexpnG{G} \log r}}$ with constant
probability.
\end{restatable}
We show in Corollary \ref{cor:alg1} that Theorem \ref{thm:alg1} implies the 
same approximation guarantee for conductance in undirected polymatroidal 
networks as well.
We also show how to extend Theorems \ref{thm:mult-approx} and \ref{thm:alg1}
to obtain analogous results for polymatroidal hypergraphs.
We prove Theorems \ref{thm:mult-approx-h} and \ref{thm:alg1-h}
in Section \ref{sec:h}.
\begin{restatable}{theorem}{multapproxh}
\label{thm:mult-approx-h}
There is a polynomial time $\bigo{\sqrt{\log n}}$-approximation algorithm for 
$\spar_H$ of a polymatroidal hypergraph
$H = \paren{V,E,(w_i)_{i \in V},(\fplus{v}{\cdot},\fminus{v}{\cdot})_{v \in V}}$.
\end{restatable}

\begin{restatable}{theorem}{talgcheegerh}
\label{thm:alg1-h}
Let $H = \paren{V,E,(w_i)_{i \in V},(\fplus{e}{\cdot},\fminus{e}{\cdot})_{e \in E}}$
be a polymatroidal hypergraph, and let
$r \defeq \max_{e \in E} \Abs{e}$.
There is a randomized polynomial time algorithm that outputs a set $S$ 
such that $\uexpn{H}{S} = \bigo{\sqrt{\uexpnG{H} \log r}}$ with constant
probability.
\end{restatable}

We show that Theorem \ref{thm:alg1-h} implies the known approximation bounds for
symmetric vertex expansion \cite{LouisRV13}; see Section \ref{sec:vert-exp}.
We also study a new notion of hypergraph conductance which we 
believe could be of independent interest; see Section \ref{sec:hyper-exp}.
We also obtain a $\bigo{\sqrt{\opt \log r}}$ approximation for this; see Theorem 
\ref{thm:alg2-h}.

Finally, we study a natural notion of an eigenvalue/spectral gap of a 
polymatroidal network, denoted by $\gamma_2$, generalizing the 
corresponding notions of graphs (folklore) and hypergraphs \cite{ChanLTZ18}. 
We prove a ``Cheeger'' inequality for polymatroidal networks relating 
the conductance and $\gamma_2$.
We define $\gamma_2$ and prove Theorem \ref{thm:cheeger} in Section
\ref{sec:cheeger}.
\begin{restatable}{theorem}{tcheeger}	
\label{thm:cheeger}
Let $G = \paren{V,E,(w_i)_{i \in V},(\rhoplussymb{v},\rhominussymb{v})_{v \in V}}$
be a polymatroidal network. Then, 
	\[ \frac{\gamma_2}{2} \leq \expnG{}{G} \leq \bigo{\sqrt{\gamma_2}} . \]
\end{restatable}

\subsection{Discussion and Related Work}
\label{sec:related}
Flows, cuts and partitioning are extensively studied in graphs and
hypergraphs under a variety of settings. The idea of combining
submodularity and graphs goes back to the classical work of Edmonds
and Giles \cite{EdmondsG77} and Hassis \cite{Hassin78} and Lawler and
Martel \cite{LawlerM82}.  There is a sense in which the polymatroidal
network model is ``local'' in that the submodularity constraints apply
only at each node as opposed to ``globally''. There is a good reason
for this. One can consider more general models that link edges of the
entire graph via a submodular function --- however this leads to
intractability with polynomial-factor hardness results even for the
$s$-$t$ cut case \cite{SvitkinaF11,JegelkaB11} while one has
maxflow-mincut equivalence in the polymatroidal network setting.  We
mention the specific result of Svitkina and Fleischer
\cite{SvitkinaF11}.  They consider the abstract problem of finding the
sparsest cut of a symmetric submodular function
$f:2^V \rightarrow \mathbb{R}_+$ defined as follows: given $f$ find
$S$ to minimize $\frac{f(S)}{|S||V-S|}$. They prove that obtaining an
$\omega(\sqrt{\log n/n})$-approximation requires an exponential number
of value queries to $f$.

As we mentioned, there has been recent interest in submodular
hypergraphs \cite{Yoshida19,LiM18,ChenOT23,Veldt23}.  Before going
into finer discussion we describe the high-level set up. The input
consists of a hypergraph $G=(V,E)$ and a submodular function
$f_e:2^e \rightarrow \mathbb{R}_+$ for each edge $e$; we do not assume
that $f_e$ is monotone. The goal is to find a sparse cut $(S,V-S)$
where the value of the cut depends on the edges $\delta_G(S)$ crossing
the cut and the their associated submodular functions.  One natural
setting, keeping in line with undirected hypergraphs, is to assume
that each $f_e$ is symmetric, and to define the cut value as
$\sum_{e \in \delta_G(S)} f_e(e \cap S) = \sum_{e \in \delta_G(S)}
f_e(e \cap (V-S))$. Even in this setting, one can easily see that if
we let $f_e$ to be an arbitrary symmetric submodular function then the
strong negative results of \cite{SvitkinaF11} already apply with a
single spanning hypergraph! Nevertheless \cite{Yoshida19,LiM18} work
in this generality (and more) and mainly prove spectral results that
do not necessarily yield polynomial-time approximation algorithms. In
some cases they yield efficient algorithms with approximation bounds
that depend on parameters of the functions in non-trivial ways. More
recent work \cite{ChenOT23,Veldt23} restrict the class of functions
and obtain polynomial-time approximation algorithms.

It may not be quite obvious how submodular hypergraphs are related to
graph cuts. We note that cuts in standard hypergraphs can be
understood via node-weighted cuts via the standard bipartite graph
representation of hypergraphs (also called factor graph representation
in \cite{ChenOT23, Veldt23}). Polymatroid cuts capture node weights
(see \cite{ChekuriKRV15} for instance) and thus it is not surprising
that we can capture classes of submodular hypergraphs via
polymatroidal networks. We explicate this in more detail later.

Now we describe the results and connections to some of the preceding papers in more detail.
Veldt \cite{Veldt23} considers sparsest cut in submodular hypergraphs
with symmetric functions $f_e$ as described above, and further
constrains each $f_e$ to be cardinality based. That is, it is assumed
that $f_e(A) = f_e(B)$ if $|A| = |B|$. Sparsest cut in this setting
can be captured by polymatrodal networks. Veldt obtains an
$O(\log n)$-approximation via the cut-matching framework
\cite{KRV,OSSV} which was refined in \cite{OrecchiaATT22}. As we
remarked previously, an $O(\log n)$-approximation for sparsest cut is
known from \cite{ChekuriKRV15}; the goal in \cite{Veldt23} was to
develop fast algorithms

Chen, Orecchia and Tani \cite{ChenOT23} considered polymatroidal cut
functions for submodular hypergraphs (see Definition \ref{def:spar-h} for formal definition) that are inspired by the model in
\cite{ChekuriKRV15} (a previous version of \cite{ChenOT23} had a
simpler model and was done concurrently and independently of
\cite{Veldt23}). They obtained an $O(\sqrt{\log n})$-approximation
for the sparsest cut problem via SDP and techniques from
\cite{AroraRV09,AgarwalCMM05}. They also obtain a fast $O(\log n)$-approximation via the
cut-matching framework. To obtain this result they designed a more
refined and stronger cut-matching framework that applied to directed
graphs and improved the $O(\log^2 n)$-bound of \cite{Louis10}. We show
that the $O(\sqrt{\log n})$-approximation for directed polymatrodial
networks can be utilized via the factor graph to derive an
approximation for the model in \cite{ChenOT23}. We do not address the
cut-matching aspects in this paper.

The work in \cite{Veldt23,ChenOT23} followed earlier work by Yoshida
\cite{Yoshida19} and Li and Milenkovic \cite{LiM18}. The focus in both
these papers is on the spectral approach and Cheeger like inequalities
for conductance. \cite{Yoshida19} studies Cheeger inequalities for
``submodular transformations''; this can essentially be viewed as
having a submodular function $f_e$ for each hyperedge $e \in E$. Their
definition of conductance of set $S$ of vertices is
$\frac{\min \set{\sum_{e \in E} f_e(e \cap S), \sum_{e \in} f_e(e \cap
    (V \setminus S))}}{ \min \set{\sum_{v \in S} \deg_v , \sum_{v \in
      V \setminus S} \deg_v }}$. Note that $f_e$ are not assumed to be
symmetric and hence the terms $\sum_{e \in E} f_e(e \cap S)$ and
$\sum_{e \in} f_e(e \cap (V \setminus S)$ can be different, however
the min essentially symmetrizes the cut definition.  For a vector
$x \in \R^n$, they define its Rayleigh quotient to be
$\frac{\sum_{e \in E} \hat{f}_e^2(x)}{\norm{x}_2^2}$; here $\hat{f}$
is the Lovasz-extension of $f_e$ and we refer the reader to
Section~\ref{sec:background}.  They define their $\lambda_2$ based on
their definition of Rayleigh quotient in the usual way and prove a
Cheeger inequality.  Their definition of conductance is similar in
spirit to ours (Definition \ref{def:expn-h}), but there are some
technical differences.  Firstly, their definition of a vertex's degree
is different from ours; they define the degree of a vertex to be the
number of hyperedges it has a non-zero contribution to.  Secondly, our
definition allows us to choose the $\min$ between $S$ and
$V \setminus S$ for each hyperedge; moreover we allow different
functions for $e \cap S$ and $e \cap (V \setminus S)$.  Finally, their
definition works for any submodular functions whereas ours only works
for monotone submodular functions.  Their definition of Rayleigh
quotient is also different from ours.  In contrast to their
definition, the numerator of our notion of Rayleigh quotient
corresponds to $\sum_{e \in E} \hat{f}_e(x^2)$.  Observe that if
$\hat{f}_e(x) > 1$, then $\hat{f}_e^2(x) \gg \hat{f}_e(x)$.  They also
give an algorithm to compute a vector $x$ whose Rayleigh quotient is
at most $O(\frac{\log r}{\e^2} \lambda_2 + \e B^2)$ where $B$ is the
maximum $\ell_2$-norm of a point in the base polytopes of the $f_e$'s.

\cite{LiM18} study submodular transformations where each $f_e$ is a 
non-negative symmetric submodular function. Some of their results
are similar to our results; their results holds for submodular transformations 
whereas ours are for directed polymatroidal networks.
Their definition of conductance of a set $S$ of vertices is $\paren{\sum_{e \in} f_e(S)}/\paren{
\min \set{\sum_{v \in S} \mu_v , \sum_{v \in V \setminus S} \mu_v }}$
where $\mu$'s are arbitrary vertex weights. Their define the Rayleigh quotient
of a vector $x$ more generally for 
any $p$-norm ($p \geq 1$) and also in terms of the vertex weights; 
their definition of the Rayleigh quotient 
for $p = 2$ is the same as ours, and hence their definition of the 
spectral gap for $p=2$ is the same as ours. They also prove a Cheeger's inequality
relating the $p$-norm spectral
gap to conductance; when $p = 2$ and the vertex weights are equal to the 
vertex degrees, their Cheeger's inequality is similar to ours (as mentioned 
above, their results holds for submodular transformations 
whereas ours are for polymatroidal networks).
Their notion of spectral gap is also not known to be efficiently computable.
They give an $r$-approximation algorithm for computing the spectral gap.
In contrast to their work, our result obtains a bound that depends on $\log r$ rather than $r$.
They also study the notion of ``nodal domains'' in submodular hypergraphs.

\paragraph{Expansion in (vanilla) hypergraphs.} 
The degree of vertex is typically defined as the number of hyperedges it 
belongs to. Recall that in Definition \ref{def:conductance}, the definition
of the conductance of a set $S$ is based on the volume of a set $S$ which 
is defined as $\vol{}{S} = \sum_{i \in S} \deg_i$. A hyperedge $e$ contributes
to the degrees of $\Abs{e}$ vertices in $S$, and therefore, the 
$\vol{}{S}$ overcounts the number of hyperedges incident on $S$. 
Note that this is not a problem in the case of graphs since each edge
contributes to the degree of only two vertices. 
Previous works have studied some approaches to 
compensate for this overcounting in hypergraphs.
Louis and Makarychev \cite{LouisM16} defined 
the expansion of a hypergraph as $\min_{S \subset V} \Abs{\delta(S)}/\Abs{S}$. They gave 
a $\bigo{\sqrt{\log n}}$-approximation algorithm and a 
$\bigo{\sqrt{(\deg_{\max}/r) \opt \log r}}$ approximation for $r$-uniform hypergraphs (here $\deg_{\max} \defeq \max_{i \in V} \deg_i$). They prove a more
general form of the latter bound which also works for non-uniform hypergraphs and gives a better bound in some cases;
we refer the reader to \cite{LouisM16} for the precise statement
(their algorithm is also for the more general problem of small-set expansion). 
\cite{Louis15,ChanLTZ18} studied conductance defined as 
$\min_{S \subset V} \Abs{\delta(S)}/\min \set{\sum_{i \in S} \deg_i,
\sum_{i \in V \setminus S} \deg_i}$. They gave a $\bigo{\sqrt{
\paren{\opt \log r}/r_{\min}}}$ approximation bound (recall that $r_{\min} \defeq \min_{e \in E} \Abs{e}$).
In Section \ref{sec:hyper-exp}, we study the conductance 
of (vanilla) hypergraphs defined as $\min_{S \subset V} \Abs{\delta(S)}/
\min \set{ \Abs{\set{e \in E : e \cap S \neq \emptyset}}, 
\Abs{\set{e \in E : e \cap (V \setminus S) \neq \emptyset}}}$
which can be viewed as more direct way of counting the number of hyperedges 
incident on $S$. 

\paragraph{Vertex expansion in graphs.} Given a graph $G = (V,E)$ with 
vertex weights $(w_i)_{i \in V}$, its vertex expansion is defined as
$\min_{S \subset V} \paren{\sum_{i \in N(S)} w_i}/\paren{\sum_{i \in S} w_i}$;
here $N(S)$ denotes the neighborhood of $S$, i.e. the set of vertices in 
$V \setminus S$ which have a neighbour in $S$. The symmetric
vertex expansion is defined as $\min_{S \subset V} 
\paren{\sum_{i \in N(S) \cup N(V \setminus S)} w_i}/
\min \set{\paren{\sum_{i \in S} w_i},\paren{\sum_{i \in V \setminus S} w_i}}$.
Feige et al. \cite{FeigeHL08} gave a $\bigo{\sqrt{\log n}}$ approximation for
vertex expansion. Louis et al. \cite{LouisRV13} give a $\bigo{\sqrt{\opt \log r}}$
approximation for vertex expansion where $r$ is the largest vertex degree.
\cite{LouisRV13} essentially showed that the approximability of vertex
expansion and symmetric vertex expansion are within a constant factors for each other.
Therefore, both the above mentioned approximation guarantees hold for
symmetric vertex expansion as well.
\cite{LouisM16} gave reduction from symmetric vertex expansion in graphs 
to hypergraph expansion; therefore these two quantities are very closely related.

\paragraph{Cheeger's inequalities.} The classical Cheeger's inequality
for graphs \cite{Alon86,AlonM85} relates $\lambda_2$, the second
smallest eigenvalue of the normalized Laplacian matrix of the graph, and
the graph's conductance. $\lambda_2$ is also equal to minimum value over
the set of vectors orthogonal to the eigenvector corresponding to the smallest
eigenvalue (the all-ones vector for regular graphs), of a quantity called
the ``Rayleigh quotient'' $\cR : \R^n \to \R$. There are no 
natural Laplacian matrices for proving 
Cheeger's inequalities for 
vertex expansion in graphs or conductance in hypergraphs. 
Therefore, prior works have produced appropriate notions of 
Rayleigh quotient for the problem being studied and defined
the notion of an ``eigenvalue'' in an analogous manner.
This type of a definition of eigenvalue has been used to prove a 
Cheeger's inequality for vertex expansion in graphs by Bobkov et al. 
\cite{Bobkovht00}.
Chan et al. \cite{ChanLTZ18} proved an analogous result for hypergraph
conductance.
As mentioned earlier in more detail, the works of Yoshida \cite{Yoshida19}
and Li and Milenkovic \cite{LiM18}, can also be viewed proving a Cheeger's
inequality for submodular transformations using this type of a definition
of eigenvalue.
Our Cheeger's inequality (Theorem \ref{thm:cheeger}) for polymatroidal networks
also uses this type of a definition of eigenvalue.
While the proofs of all these results above may seem to be following
the same template, there are subtle technical challenges which require
new ideas to overcome them. 

While the eigenvalue in the case a graph conductance is computable since it 
is equal to an eigenvalue of the normalized Laplacian matrix of the graph, 
the eigenvalues in the other cases mentioned above are not known to be 
efficiently computable in general. There have been works studying approximation 
algorithms for computing these eigenvalues. In the case of vertex expansion in graphs 
\cite{SteurerT12,LouisRV13} gave an $\bigo{\log r}$ approximation algorithm
for computing the eigenvalue, where $r$ is the largest vertex degree. 
For hypergraph conductance \cite{ChanLTZ18} gave an $\bigo{\log r}$ 
approximation algorithm for computing the eigenvalue, where $r$ is the
cardinality of the largest hyperedge. 
The approximations for the case of submodular transformations \cite{Yoshida19,LiM18}
have been mentioned earlier.
The proof of Theorem \ref{thm:alg1} can also be viewed as giving 
an $\bigo{\log r}$  approximation algorithm to the eigenvalue in the case
of symmetric conductance in polymatroidal networks.

We also mention the recent works of Kwok et al. \cite{KwokLT22} which
proved a Cheeger's inequality for vertex expansion, and Lau et al. 
\cite{LauTW23} which proved Cheeger's inequalities for hypergraph 
expansion using the notion of ``reweighted eigenvalues''.
In the case of vertex expansion in graphs, the reweighted eigenvalue
refers to largest value that can be attained by the second smallest 
eigenvalue of the normalized Laplacian matrix of graph by reweighting
the edges of graph such that the weighted degree of each vertex is 
one. We refer the reader to these papers for a more detailed discussion.

\section{Preliminaries and Background}
\label{sec:background}

Most of this material is taken directly from \cite{ChekuriKRV15} and
we include it for the sake of completeness. A reader familiar with the
material can skip relevant parts.

\subsection{Submodular functions and continuous extensions}
\label{subsec:submod-extensions}

\paragraph{Lov\'{a}sz extension:} For a set function $\rho:2^N \to
\R$ (not necessarily submodular) its Lov\'{a}sz extension
\cite{Lovasz83} denoted by $\hrhosymb:[0,1]^N \to \mathbb{R}$
is defined as follows:
\begin{equation}
\label{eq:lovext}
\hrho{}{x} = \int_0^1 \rho\paren{\set{i \mid x_i \geq \theta}} d\theta,
\end{equation}
This is not the standard
way the Lov\'{a}sz extension is stated but is equivalent to it.
The standard definition is the
following. Given $x$ let $i_1,\ldots,i_n$ be a permutation of
$\{1,2,\ldots,n\}$ such that $x_{i_1} \ge x_{i_2} \ge \ldots \ge x_{i_n} \ge 0$.
For ease of
notation define $x_0=1$ and $x_{n+1} = 0$.
For $1 \le j \le n$ let $S_j = \{i_1,i_2,\ldots,i_j\}$.
Then
$$ \hrho{}{x} = (1-x_{i_1})\rho(\emptyset) + \sum_{j=1}^n (x_{i_j}-x_{i_{j+1}})\rho(S_j).$$
It is typical to assume that $\rho(\emptyset) = 0$ and omit the first
term in the right hand side of the preceding equation. Note that it is
easy to evaluate $\hrho{}{x}$ given a value oracle for $\rho$.

We state some well-known facts.

\begin{fact}[\cite{Lovasz83}]
\label{fact:convexity}
The Lov\'{a}sz extension of a submodular function is convex.
\end{fact}

\begin{fact}
  \label{fact:monotone-sub}
  For a monotone submodular function $\rho$ and $x \le x'$
	(coordinate-wise), $\hrho{}{x} \le \hrho{}{x'}$.
\end{fact}

\paragraph{Base polytope and relation to extensions:} Let
$\rho:2^N \to \Rgeq$ be a normalized monotone
submodular function. The base polytope of $\rho$, denoted by
$\basepoly{\rho}$, is the polytope in $\mathbb{R}^N$ given by the
following inequalities:
\begin{eqnarray}
  \label{eq:basepolytope}
  \sum_{i \in S} x_i & \le & \rho(S) \quad \forall S \subseteq N \\
  \sum_{i \in N} x_i & = & \rho(N) \\
  x_i & \ge & 0 \quad \forall i \in N
\end{eqnarray}

Given a weight vector $w \in \mathbb{R}^n$ the problem $\max_{x
\in \basepoly{\rho}} w^T x$ can be solved via a simple greedy algorithm as shown in the
classical work of Edmonds (see ~\cite{Schrijver-book}). Further, the
following is also known.

\begin{lemma}
\label{lem:lov-ext}
For a monotone submodular set function $\rho: 2^N \to
	\Rgeq$ and for any $x \in  [0,1]^N$, $\hrho{}{x} =
	\max_{y \in \basepoly{\rho}} x^T y$. 
\end{lemma}

\begin{lemma}
\label{lem:summ}
Let $\hat{\rho} : \R^N \to \R$ be the \lovasz~ extension of a monotone submodular
function $\rho : \set{0, 1}^N \to \R$. 
Let $x, x_1, x_2 \in \Rgeq^N$ such that $x \leq x_1 + x_2$. Then
$\hat{\rho}(x) \leq \hat{\rho}(x_1) + \hat{\rho}(x_2)$.
\end{lemma}

\begin{proof}
Fix $x,x_1,x_2$ as in the statement of the lemma.
Let $\basepoly{\rho}$ be the base polytope of $\rho$. Then using Lemma \ref{lem:lov-ext}, 
$\hat{\rho}(x) = \max_{w \in \basepoly{\rho}} \inprod{w,x}$.
Since $\rho$ is a monotone submodular function and $x, x_1, x_2 \geq 0$, 
there is an optimal $w$ with only non-negative
values. Therefore,
\begin{align*}
\hat{\rho}(x) & = \max_{w \in \basepoly{\rho}} \inprod{w,x} 
	= \max_{\substack{w \in \basepoly{\rho} \\ w \geq 0}} \inprod{w,x} 
	 \leq \max_{\substack{w \in \basepoly{\rho} \\ w \geq 0}} \inprod{w,x_1 + x_2} 
	& \textrm{(Using $x, x_1, x_2 \geq 0$)} \\
	& \leq \max_{\substack{w \in \basepoly{\rho} \\ w \geq 0}} \inprod{w,x_1} 
	+ \max_{\substack{w \in \basepoly{\rho} \\ w \geq 0}} \inprod{w,x_2} \\
	& = \hat{\rho}(x_1) + \hat{\rho}(x_2) .
\end{align*}
\end{proof}

\begin{fact}
\label{fact:ax}
For a monotone submodular function $\rho: 2^N \to \Rgeq$, we have
\[ \hrho{}{ax} = a \hrho{}{x} \qquad \forall x \in \Rgeq^N, \ a \in \Rgeq . \]
\end{fact}

\begin{proof}
Let $\basepoly{\rho}$ be the base polytope of $\rho$.
Fix $x \in \Rgeq^N, \ a \in \Rgeq$.
\[ \hrho{}{ax} = \max_{w \in \basepoly{\rho}} w^T (ax) =  a \max_{w \in \basepoly{\rho}} w^T x = a \hrho{}{x} . \]
\end{proof}

\subsection{Cuts and flows in polymatroidal networks}
\label{sec:polymatroidal}
We briefly discuss (multicommodity) cuts and flow problems in polymatroidal
networks with the goal of explaining the cost of the cut in these
networks. In addition to the graph, the input consists of a set of $k$
source-sink pairs $(s_1,t_1),\ldots,(s_k,t_k)$, and in some cases
additional demand information.

\subsubsection{Cuts \label{sec:cuts}}
Given a graph $G=(V,E)$ and a set of edges $F
\subseteq E$ we say that the ordered node pair $(s,t)$ is separated by
$F$ if there is no path from $s$ to $t$ in the graph $G[E\setminus
F]$.  In the standard network model the cost of a cut
defined by a set of edges $F$ is simply $\sum_{e \in F} c(e)$ where
$c(e)$ is the cost of $e$ (capacity in the primal flow network) . In
polymatroid networks the cost of $F$ is defined in a more involved
fashion. Each edge $(u,v)$ in $F$ is assigned to either $u$ or $v$; we
say that an assignment of edges to nodes $g_F: F \to V$ is {\em
  valid} if it satisfies this restriction.  A valid assignment
partitions $F$ into sets $\{ g_F^{-1}(v) \mid v \in V\}$ where
$g_F^{-1}(v)$ (the pre-image of $v$) is the set of edges in $F$ assigned
to $v$ by $g_F$.  For a given valid assignment $g_F$ of $F$ the cost of
the cut $\cut{G}{F,g_F}$ is defined as
\[ \cut{G}{F,g_F} \defeq \sum_{v \in V} \paren{\rhoplus{v}{\delta^+(v) \cap g_F^{-1}(v)} 
	+ \rhominus{v}{\delta^-(v) \cap g_F^{-1}(v)}}.\]

Given a set of edges $F$ we define its cost to be the minimum over
all possible valid assignments of $F$ to nodes, the expression for the
cost as above. We give a formal definition below.

\begin{definition}[Cost of cut]
\label{defn:cut-cost}
Given a directed polymatroid network 
$G = \paren{V,E,(w_i)_{i \in V},(\rhoplus{v}{\cdot},\rhominus{v}{\cdot})_{v \in V}}$
	and a set of edges $F \subseteq E$, its cost denoted by $\cut{G}{F}$ is
\[ \cut{G}{F} \defeq \min_{g \in \set{F \to V}} \cut{G}{F,g} . \]
In an undirected polymatroid network $\cut{G}{F}$ is
\[ \cut{G}{F} \defeq \min_{g: F \to V} \sum_{v \in V} \rho_v(g^{-1}(v)).
\]
\end{definition}
Given a collection of source-sink pairs $(s_1,t_1),\ldots,(s_k,t_k)$
in $G$ and associated demand values $D_1,\ldots,D_k$, and a
set of edges $F \subseteq E$, the demand separated by $F$, denoted by
$D(F)$, is $\sum_{i:(s_i,t_i) \text{~separated by~} F} D_i$. $F$ is
a {\em multicut} if all the given source-sink pairs are separated by $F$. The
{\em sparsity} of $F$ is defined as $\frac{\cut{G}{F}}{D(F)}$.

Given the above definitions two natural optimization problems that arise are
the following. The first is to find a multicut of minimum cost for a
given collection of source-sink pairs. The second is to find a cut of
minimum sparsity. We say that an instance of sparsest cut is a
product-form or uniform sparsest cut if there are non-negative weights
$(w_i)_{i \in V}$, and $D_{u,v} = w_u w_v$ for
each pair of vertices $u,v \in V$. In this case, the above definition of 
sparsity reduces to Definition \ref{def:sparsity}.

\paragraph{Conductance:} Conductance in the setting of graphs can be
viewed as a special case of sparsest cut for product multicommodity
flow where $w_u = \deg_u$ under the assumption that the graph is
unweighted --- otherwise we define it as the weighted degree. This
particular weight function has important connections to random walks
and spectral theory via the well-known Cheeger's inequality. The
definition of degree becomes less clear when dealing with more general
settings, especially in the context of polymatroidal networks and (submodular) 
hypergraphs.  For our purposes here we define the degree of a vertex $v$ to be
$\deg_v \defeq \rhoplus{v}{\delta^+(v)} + \rhominus{v}{\delta^-(v)}$. Note that if $G$
is bidirected graph with equal capacities on each of the directions of
an edge, the degree corresponds to twice the usual degree. We use $D$
to denote the $n \times n$ diagonal matrix where $D_{ii} = \deg_i$ and
$W$ to denote the $n \times n$ diagonal matrix where $W_{ii} = w_i$.

\subsubsection{Flows \label{sec:flows}}
A multicommodity flow for a given collection of $k$ source-sink pairs
$(s_1,t_1),\ldots,$ $(s_k,t_k)$ consists of $k$ separate single-commodity
flows, one for each pair $(s_i,t_i)$. The flow for the $i$'th commodity
can either be viewed as an edge-based flow $f_i:E \to \Rgeq$,
or as a path-based flow $f_i: \cP_i \to \Rgeq$, where
$\cP_i$ is the set of all simple paths between $s_i$ and $t_i$ in $G$.
Given path-based flows $f_i$, $i=1,\ldots,k$ for the $k$ source-sink pairs,
the total flow on an edge $e$ is defined as
$f(e) = \sum_{i=1}^k \sum_{\substack{p \in \cP_i:\\e \ni p}} f_i(p)$.
The total flow for commodity $i$ is
$F_i  =  \sum_{p \in \cP_i} f_i(p)$.
In directed polymatroidal networks, the flow is constrained to satisfy
the following capacity constraints.
\[ \sum_{e \in S} f({e}) \leq \rhominus{v}{S} \quad  \forall S \subseteq \delta^-(v)
  \qquad \textrm{and} \qquad
\sum_{e \in S} f({e})  \leq \rhoplus{v}{S} \quad \forall S \subseteq
\delta^+(v) . \]

\medskip Given rational valued demand tuple
$\mathcal{D} = (D_1,\ldots,D_k)$ where $D_i$ is the demand for pair
$(s_i,t_i)$, we say $\mathcal{D}$ is {\em achievable} if there is a
feasible multicommodity flow in $G$ that routes $D_i$ flow for pair
$i$.  For a given polymatroidal network $G$ and given set of
source-sink pairs $\cT$, the set of achievable demand tuples can be
seen to be a polyhedral set.  We let $P(G,\cT)$ denote this region.
In the {\em maximum throughput multicommodity flow} problem the goal
is to maximize $\sum_{i=1}^k D_i$ over $P(G,\cT)$.  In the {\em
  maximum concurrent multicommodity flow} problem each source-sink
pair has a given demand $D_i$, and the goal is to maximize $\lambda$
such that the tuple $(\lambda D_1,...,\lambda D_k)$ is achievable,
that is the tuple belongs to $P(G,\cT)$. It is easy to see that both
these problems can be cast as linear programming problems. The
path-formulation results in an exponential (in $n$ the number of nodes
of $G$) number of variables and we also have an exponential number of
constraints due to the polymatroid constraints at each node. However,
one can use an edge-based formulation and solve the linear programs in
polynomial time via the ellipsoid method and polynomial-time
algorithms for submodular function minimization.

\subsection{Relaxations for sparsest cut and conductance}
\label{sec:sparsest-cut-relaxation}
In this section we describe relaxations for computing the sparsest
cut in polymatroidal networks. These mimic the relaxations for
sparsest cut in graphs that have been extensively studied.
LP relaxatiosn for polymatroidal networks were studied in \cite{ChekuriKRV15}
where the focus was establishing flow-cut gaps. It was suggested in
\cite{ChekuriKRV15} that certain results from graphs via SDP
relaxations can be ported over to polymatroidal networks based on the
line embedding based approach developed in that paper. 
We first describe the easier to understand
LP based relaxation studied in \cite{ChekuriKRV15} and then
describe SDP based relaxations. Later we will also spectral
relaxations.

Consider the cost of a cut and how to model it in a fractional way. 
Note that $\cut{}{F}$ is defined by
valid assignments of $F$ to the nodes, and submodular costs on the
nodes. In the relaxation we model this as follows. For an edge
$e=(u,v)$ we have variables $\ell(e,u)$ and $\ell(e,v)$ which decide
whether $e$ is assigned to $u$ or $v$.  We have a constraint
$\ell(e,u) + \ell(e,v) = \ell(e)$ to model the fact that if $e$ is cut
then it has to be assigned to either $u$ or $v$.  Now consider a node
$v$ and the edges in $\delta^+(v)$.  The variables $\ell(e,v), e \in
\delta^+(v)$ in the integer case give the set of edges $S \subseteq
\delta^+(v)$ that are assigned to $v$ and in that case we can use the
function $\rho^+_v(S)$ to model the cost.  However, in the fractional
setting the variables lie in the real interval $[0,1]$; we use
the Lov\'{a}sz extension to obtain a linear program.
Let $\dv{-}{\ell}{}{v}$ be the vector consisting of the
variables $\ell(e,v)$, $e \in \delta^-(v)$ and similarly $\dv{+}{\ell}{}{v}$
denote the vector of variables $\ell(e,v)$, $e \in \delta^+(v)$.

In the sparsest cut problem we need to decide which pairs to
disconnect and then ensure that we pick edges whose removal separates
the chosen pairs. Moreover we are interested in the ratio of the cost
of the cut to the demand separated.  We follow the standard
formulation in the edge-capacitated case with the main difference,
again, being in the cost of the cut. There is a variable $y_i$ which
determines whether pair $i$ is separated or not. We have the
edge variables $\ell(e), \ell(e,u), \ell(e,v)$ to indicate whether
$e=(u,v)$ is cut and whether $e$'s cost is assigned to $u$ or $v$. If
pair $i$ is to be separated to the extent of $y_i$ we need ensure that
the $\ell(e)$ variables induce a distance of at least $y_i$. For this
purpose we use $\ell(s_i,t_i)$ as the shortest path distance
between
$s_i$ and $t_i$ induced by lengths on edges given by $\ell: E
\to \Rgeq$. We then add the constraint that
$\ell(s_i,t_i) \ge y_i$. Note that this constraint can be
implemented via a set of exponential linear constraints, one for each
path between $s_i$ and $t_i$, or can be implemented via a cubic number
of triangle inequality constraints. Since this is standard we do not
explicitly write down the constraints that implement this ``meta'' constraint.
To express sparsity, which is defined
as a ratio, we normalize the demand separated to be $1$.
The relaxation for the undirected uniform case is shown
in Fig~\ref{fig:sparsestcut-relaxations} where
$\dv{}{\ell}{}{v}$ is the vector of variables $\ell(e,v), e \in \delta(v)$; the $y_i$ variables can be completely eliminated from the LP using standard techniques.

\begin{figure}[htb]
  \centering
  \begin{boxedminipage}[t]{0.45\linewidth}
    \begin{align*}
	\min \sum_v \hrhominus{v}{\dv{-}{\ell}{}{v}} & + \hrhoplus{v}{\dv{+}{\ell}{}{v}} \\
  \ell(e,u) + \ell(e,v) & =  \ell(e) \quad \quad  e = (u,v) \in E \\
  \sum_{i,j \in V} w_i w_j \ell(i,j) & =  1 \\
  \ell(e), \ell(e,u),\ell(e,v) & \ge  0 \quad \quad e = (u,v) \in E.
    \end{align*}
\end{boxedminipage}
  \caption{LP relaxation for uniform sparsest cut in a directed polymatroidal network}
  \label{fig:sparsestcut-relaxations}
\end{figure}

One can show that the relaxations for the sparsest cut shown in
Fig~\ref{fig:sparsestcut-relaxations} are dual to the ``maximum
concurrent flow'' in the primal; we refer the reader to \cite{ChekuriKRV15}.

\paragraph{SDP Relaxations:} We extend the ideas behind the LP relaxation to 
develop an SDP relaxations via a vector based formulation. 
To model the variables such as $\ell(e,u)$ in the LP formulation, the vector formulation 
embeds both edges and vertices of the graph as vectors in (high-dimensional) Euclidean space. 
We have a vector variable $\vect{v}$ for each vertex $v \in V$ and a vector variable
$\vect{e}$ for each edge $e \in E$. Since we are working with directed works, we need the ability to model 
asymmetric distances that are induced by these vectors. Here we borrow basic ideas from \cite{AgarwalCMM05}.
The formal SDP relaxation to obtain an $O(\sqrt{\log n})$ approximation is described and analyzed
in Section~\ref{sec:mult-approx}. To obtain a spectral bound for (symmetric) conductance, the SDP can be simplified and
we describe it in Section~\ref{sec:alg1}. 
For proving Cheerger-like inequalities (Theorem \ref{thm:cheeger}), one
can also work with Rayleigh coefficients that are not efficiently
computable but are of interest from a mathematical point of view.

\paragraph{Efficiently solving the SDP.}
Using Fact \ref{fact:convexity} we get that the SDP relaxations
in this paper are convex programs. Therefore, they can be solved
efficiently using standard algorithms \cite{Schrijver-book}.

\subsection{Line embeddings}
In the context of sparsest cut for edge-capacitated undirected graphs,
it is known that $\ell_1$-embeddings of metrics capture the tight
flow-cut gap \cite{LLR95,AR98,GNRS04}. Line embeddings with average
distortion \cite{Rabinovich08} suffice in many settings and they are particularly useful
for vertex-capacities \cite{FeigeHL08}, hypergraphs \cite{ChanLTZ18},
polymatroidal networks \cite{ChekuriKRV15} and Cheeger-like
inequalities in the classical setting as well as in recent works on
submodular transformations \cite{Yoshida19,LiM18}.
Our algorithms will also proceed by producing a good line embedding of
polymatroidal graph.

We now discuss what we mean by a ``good'' line embedding.
Note that we are dealing with directed graphs. Due to the nature of
the cuts in polymatroidal networks, for our purposes, we consider
a line embedding as a mapping $x: V \cup E \rightarrow \mathbb{R}$.
For a line embedding $x$ sorted in decreasing order, we will see that the out-edges that lie to the right of any vertex 
and the in-edges that lie to the left of a vertex play an important 
role in the value of $\cut{}{\delta^+(S)}$.
Similarly, the in-edges that lie to the right of any vertex 
and the out-edges that lie to the left of a vertex play an important 
role in the value of $\cut{}{\delta^-(S)}$. 
To capture this, we make the following definitions. 
We will be interested in different non-negative real-valued functions $\ell(\cdot)$ that map
scalars, vectors or pairs of vectors to reals. 
Given $\set{x_v}_{v \in V} \cup \set{x_e}_{e \in E} \subset \R$ 
and a function $\ell : \R \rightarrow \R_+$, we define 
$\dvright{+}{\ell}{x}{v}$ to be the vector of $\ell\paren{\max \set{0, x_v - x_e}}$
values for all $e \in \delta^+(v)$,
$\dvright{-}{\ell}{x}{v}$ to be the vector of $\ell\paren{\max \set{0, x_v - x_e}}$
values for all $e \in \delta^-(v)$,
$\dvleft{+}{\ell}{x}{v}$ to be the vector of $\ell \paren{\max \set{0,x_e - x_v}}$
values for all $e \in \delta^+(v)$,
and 
$\dvleft{-}{\ell}{x}{v}$ to be the vector of $\ell \paren{\max \set{0,x_e - x_v}}$
values for all $e \in \delta^-(v)$.
Typically, $\ell(x)$ will be $\norm{x}_1$, i.e. the $\ell_1$ norm, or
$\norm{x}_2^2$, i.e. the  $\ell_2^2$ value (note that in general this
is not a norm).

The results in \cite{ChekuriKRV15} were based on the interplay between
the Lov\'asz-extension and line embeddings. We will use the following lemma from \cite{ChekuriKRV15} (See
Lemma 9, Lemma 10 and Remark 3 therein); we reproduce their proof in
Section \ref{sec:line-embedding} for the sake of completeness.

\begin{restatable}{lemma}{lineembedding}[\cite{ChekuriKRV15}]
\label{lem:line-embedding}
Let $\paren{V,E,(w_i)_{i \in V},(\rhoplus{v}{\cdot},\rhominus{v}{\cdot})_{v \in V}}$
be a polymatroidal network, 
and let $x \in \R^{V \cup E}_{\geq 0}$. 
Then, there is a polynomial time algorithm to
compute sets $S, S' \subseteq \supp{x}$, $g^+: \delta^+(S) \to V$
and $g^-: \delta^-(S') \to V$ satisfying
\[ \frac{\cut{}{\delta^+(S),g^+}}{\wt{}{S}} \leq 
	\frac{\sum_{v \in V} \paren{\hrhoplus{v}{\dvright{+}{\ell_1}{x}{v}} 
	+ \hrhominus{v}{\dvleft{-}{\ell_1}{x}{v}}}}{\sum_{v \in V} w_v x_v } \]
and
\[ \frac{\cut{}{\delta^-(S'),g^-}}{\wt{}{S'}} \leq 
	\frac{\sum_{v \in V} \paren{\hrhoplus{v}{\dvleft{+}{\ell_1}{x}{v}} 
	+ \hrhominus{v}{\dvright{-}{\ell_1}{x}{v}}}}{\sum_{v \in V} w_v x_v } . \]
Moreover, if some $(u,v) \in E$ is cut by $S$ and $x_u = x_{(u,v)}$, 
then $g^+((u,v)) = v$.
Similarly, if some $(u,v) \in E$ is cut by $S'$ and $x_v = x_{(u,v)}$, 
then $g^-((u,v)) = u$.
\end{restatable}

\begin{lemma}[\cite{ChekuriKRV15}]
\label{lem:line-embedding-2}
Let $\paren{V,E,(w_i)_{i \in V},(\rhoplus{v}{\cdot},\rhominus{v}{\cdot})_{v \in V}}$
be a polymatroidal network, 
and let $x \in \R^{V \cup E}$. 
Then, there is a polynomial time algorithm to
compute sets $S, S' \subseteq \supp{x}$, $g^+: \delta^+(S) \to V$
and $g^-: \delta^-(S') \to V$ satisfying
\[ \frac{\cut{}{\delta^+(S),g^+}}{\wt{}{S} \ \wt{}{V \setminus S}} \leq 
	\frac{\sum_{v \in V} \paren{\hrhoplus{v}{\dvright{+}{\ell_1}{x}{v}} 
	+ \hrhominus{v}{\dvleft{-}{\ell_1}{x}{v}}}}{\sum_{i,j \in V} w_i w_j \Abs{x_i - x_j}} \]
and
\[ \frac{\cut{}{\delta^-(S'),g^-}}{\wt{}{S'} \ \wt{}{V \setminus S'}} \leq 
	\frac{\sum_{v \in V} \paren{\hrhoplus{v}{\dvleft{+}{\ell_1}{x}{v}} 
	+ \hrhominus{v}{\dvright{-}{\ell_1}{x}{v}}}}{\sum_{i,j \in V} w_i w_j \Abs{x_i - x_j}} . \]

\end{lemma}

We will need the following ``undirected version'' of the Lemma \ref{lem:line-embedding}.
This lemma is also from \cite{ChekuriKRV15}, we reproduce their proof here for completeness.
\begin{restatable}{lemma}{lineembeddingu}[\cite{ChekuriKRV15}]
\label{lem:line-embedding-u}
Let $\paren{V,E,(w_i)_{i \in V},(\rhoplus{v}{\cdot},\rhominus{v}{\cdot})_{v \in V}}$
be a polymatroidal network, 
and let $x \in \R^{V \cup E}_{\geq 0}$. 
Then, there is a polynomial time algorithm to
compute sets $S \subseteq \supp{x}$, $g: \delta^+(S) \cup \delta^-(S) \to V$
satisfying
\[ \frac{\cut{}{\delta^+(S),g} + \cut{}{\delta^-(S),g}}{\vol{}{S}} \leq 
	2 \frac{\sum_{v \in V} \paren{\hrhoplus{v}{\dv{+}{\ell_1}{x}{v}} 
	+ \hrhominus{v}{\dv{-}{\ell_1}{x}{v}}}}{\sum_{v \in V} \deg_v x_v} . \]
\end{restatable}

\section{Sparsest Cut in Directed Polymatroidal Networks}
\label{sec:mult-approx}

In this section, we will prove Theorem \ref{thm:mult-approx}.
\multapprox*

\subsection{SDP Relaxation}
We develop an SDP relaxation of the problem via a vector formulation. As mentioned earlier,
the SDP is based on embedding both the vertices and edges of the given directed graph $G=(V,E)$ in 
Euclidean space. In addition we will have a reference vector $\vectone$ to model directed distances.
These vectors will be required to satisfy the  $\ell_2^2$ triangle inequality for all triples which
can be written as a linear constraint in the squares of the vector lengths. Given a set of vectors $(\vect{i})_{i \in V \cup E} \cup \set{\vectone}$ 
satisfying the $\ell_2^2$ triangle
inequality, the ``directed distance'' between $i,j \in V \cup E$ is defined as
\[ \ddist{i,j} \defeq \frac{1}{2}\paren{
	\norm{\vect{i} - \vect{j}}^2 - \norm{\vectone - \vect{i}}^2
	+ \norm{\vectone - \vect{j}}^2} . \]

It is known that the $\ddist{.}$ is a directed semi-metric. 
\begin{fact}[\cite{AgarwalCMM05,CharikarMM06}]
\label{fact:acmm05}
$\ddist{.}$ defined above is a directed semimetric.
\end{fact}

Recall from Section~\ref{sec:sparsest-cut-relaxation} that for each edge $(v,u) \in \delta^+(v)$ we need
a way to assign a scalar length $\ell(v,(v,u))$ to formulate the objective based on the \lovasz~ extension.
In the context of the SDP relaxation this length will be $\ddist{v,(v,u)}$. Similarly for an incoming edge
$(u,v) \in \delta^-(v)$, $\ddist{(u,v),v}$ models the length $\ell((u,v),v)$. 
We let $\dv{+}{\ddistsymb}{\vect{}}{v}$ denote the vector of $\ddist{v,(v,u)}$
for all $(v,u) \in \delta^+(\set{v})$, and $\dv{-}{\ddistsymb}{\vect{}}{v}$ denotes
the vector of $\ddist{(u,v),v}$ for all $(u,v) \in \delta^-(\set{v})$.

This leads to the SDP relaxation \ref{sdp:spar} for $\spar_G$. A similar SDP relaxation was 
studied in \cite{ChenOT23} for the sparsest cut in polymatroidal hypergraphs.

\begin{figure}[htb]
\centering
\begin{boxedminipage}[htb]{0.8\linewidth}
\begin{sdp}
	\[ \min \sum_{v \in V_G} \paren{\hrhoplus{v}{\dv{+}{\ddistsymb}{\vect{}}{v}} 
	+ \hrhominus{v}{\dv{-}{\ddistsymb}{\vect{}}{v}}}. \]
	subject to:
\begin{align}
	\sum_{i,j \in V} w_i w_j \norm{\vect{i} - \vect{j}}^2 & = 1 \label{eq:normalization}\\
	\norm{\vect{i} - \vect{j}}^2 + \norm{\vect{j} - \vect{k}}^2
		& \geq \norm{\vect{i} - \vect{k}}^2 & \forall i,j,k \in V_G \cup E_G \cup \set{\emptyset}
		\label{eq:l22}
\end{align}
\label{sdp:spar}
\end{sdp}
\end{boxedminipage}
\caption{SDP Relaxation for uniform sparsest cut in a directed polymatroidal network.}
\end{figure}

We encapsulate the fact that SDP \ref{sdp:spar} is indeed a relaxation
in the next claim. The proof is in Section \ref{sec:sdpleqspar}.

\begin{restatable}{claim}{SDPleqspar}
\label{claim:sdpleqspar}
For SDP \ref{sdp:spar}, $\sdpval \leq \spar_G$.
\end{restatable}

\subsection{Rounding the SDP solution}
We will use the following theorem from \cite{AgarwalCMM05} (see also \cite{ChenOT23}).
\begin{theorem}[\cite{AgarwalCMM05,ChenOT23}]
\label{thm:acmm}
Given a feasible solution $(v_i)_{i \in V \cup E}$ to SDP \ref{sdp:spar}
there exists a randomized polynomial-time algorithm to compute a map 
$\chi : \set{v_i}_{i \in V \cup E} \to \R$
such that the following hold with constant probability.
\begin{enumerate}
\item $\max \set{0, \chi(i) - \chi(j)} \leq \norm{\vect{i} - \vect{j}}^2
	- \norm{\vect{i}}^2 + \norm{\vect{j}}^2 \ \forall i,j \in V \cup E$.
\item $\sum_{i,j \in V} w_i w_j \Abs{\chi(i) - \chi(j)} \geq 
	\Omega\paren{\frac{1}{\sqrt{\log k}}}
		\sum_{i,j \in V} w_i w_j \norm{\vect{i} - \vect{j}}^2$,
where $k = \Abs{\set{i \in V : w_i > 0}}$. 
\end{enumerate}
\end{theorem}

\begin{algorithm}
\caption{ARV rounding algorithm}
\label{alg:arv}
\begin{enumerate}
\item Solve SDP \ref{sdp:spar} to obtain vectors 
	$(\vect{i})_{i \in V \cup E} \subset \R^{V \cup E}$.
\item Using the mapping $\chi$ from Theorem \ref{thm:acmm} define $x_i \defeq \chi(i)$
	for $i \in V \cup E$.
\item Compute a set $S$ from $x$ using the algorithm from Lemma \ref{lem:line-embedding}.
\item Output $S$.
\end{enumerate}
\end{algorithm}

\begin{proof}[Proof of Theorem \ref{thm:mult-approx}]
Using Theorem \ref{thm:acmm} in each coordinate imples that 
\[ \dvright{+}{\ell_1}{x}{v} = \paren{\max\set{0, x_v - x_e}}_{e \in \delta^+(v)}
	\leq \paren{\ddist{\vect{v}, \vect{e}}}_{e \in \delta^+(v)}
	= \dv{+}{\ddistsymb}{\vect{}}{v} \qquad \forall v \in V \textrm{ and} \]
\[ \dvleft{-}{\ell_1}{x}{v} = \paren{\max\set{0, x_e - x_v}}_{e \in \delta^-(v)}
	\leq \paren{\ddist{\vect{e}, \vect{v}}}_{e \in \delta^-(v)}
	= \dv{-}{\ddistsymb}{\vect{}}{v} \qquad \forall v \in V . \]
Fact \ref{fact:monotone-sub} implies
\[
\sum_{v \in V_G} \paren{\hrhoplus{v}{\dvright{+}{\ell_1}{x}{v}} 
	+ \hrhominus{v}{\dvleft{-}{\ell_1}{x}{v}}}
\leq \sum_{v \in V_G} \paren{\hrhoplus{v}{\dv{+}{\ddistsymb}{\vect{}}{v}} 
	+ \hrhominus{v}{\dv{-}{\ddistsymb}{\vect{}}{v}}}. \]
Using this, Theorem \ref{thm:acmm} and $k \leq n$, we get that with constant probability,
\begin{align*}
\frac{\sum_{v \in V} \paren{\hrhoplus{v}{\dvright{+}{\ell_1}{x}{v}} 
	+ \hrhominus{v}{\dvleft{-}{\ell_1}{x}{v}}}}
	{\sum_{i,j \in V} w_i w_j \Abs{x_i - x_j} }
& \leq \bigo{\sqrt{\log n}} \frac{
\sum_{v \in V} \paren{\hrhoplus{v}{\dv{+}{\ddistsymb}{\vect{}}{v}} 
	+ \hrhominus{v}{\dv{-}{\ddistsymb}{\vect{}}{v}}}}
{\sum_{i,j \in V} w_i w_j \norm{\vect{i} - \vect{j}}^2} \\
& = \bigo{\sqrt{\log n}} \cdot \sdpval .
\end{align*}

Therefore, using the algorithm from Lemma \ref{lem:line-embedding-2} on the vector $x$
finishes the proof of this theorem. 

\end{proof}

\section{Symmetric Conductance in Directed Polymatroidal Networks}
\label{sec:alg1}
In this section we prove Theorem \ref{thm:alg1}.
\talgcheeger*

\begin{figure}[thb]
\centering
\begin{boxedminipage}[htb]{0.8\linewidth}
\begin{sdp}
\label{sdp:u}
	\[ \min \sum_{v \in V} \paren{\hrhoplus{v}{\dv{+}{\ell_2^2}{\vect{}}{v}} 
	+ \hrhominus{v}{\dv{-}{\ell_2^2}{\vect{}}{v}}}. \]
	subject to:
\begin{align}
	\sum_{i \in V} \deg_i \vect{i} & = 0 \label{eq:c3-u} \\
	\sum_{i \in V} \deg_i \norm{\vect{i}}^2 & = 1 \label{eq:normalization-u}	
\end{align}
\end{sdp}
\end{boxedminipage}
  \caption{SDP Relaxation for symmetric conductance in directed polymatroidal
    networks.}
\end{figure}

The proof of Theorem \ref{thm:alg1} 
is based on rounding a solution to the SDP relaxation \ref{sdp:u}.
The relaxation also has vectors for vertices and edges of $G$ but
is simpler since we are working with the symmetric conductance and
are only interested in a certain type of guarantee. 
Constraints \ref{eq:c3-u} and \ref{eq:normalization-u} in SDP \ref{sdp:u} are 
the standard constraints used for approximating conductance in graphs (folklore),
hypergraphs \cite{ChanLTZ18}, etc. 
Let us compare SDP \ref{sdp:spar} and SDP \ref{sdp:u}. The objective function of 
the two SDPs are based on the \lovasz~ extension based framework. 
Since \ref{sdp:spar} is used to approximate directed 
sparsest cut, its objective function uses $\ddistsymb$. Since SDP \ref{sdp:u}
is used to approximate symmetric conductance, its objective function 
uses the square of the Euclidean distance. To make the analogy to the LP 
variables from Section~\ref{sec:sparsest-cut-relaxation}, we are modeling
$\ell(v,(v,u))$ by $\norm{\vect{v}-\vect{(v,u)}}^2$ and 
$\ell((u,v),v)$ by $\norm{\vect{v}-\vect{(u,v)}}^2$. 
Constraint \ref{eq:normalization} 
in SDP \ref{sdp:spar} can be written as 
\begin{align*} 
1 & = \frac{1}{2} \sum_{i \in V} \sum_{j \in V} w_i w_j \norm{\vect{i} - \vect{j}}^2
     = \frac{1}{2} \sum_{i \in V} \sum_{j \in V} w_i w_j \paren{ \norm{\vect{i}}^2 
     + \norm{\vect{j}}^2 - 2 \inprod{\vect{i},\vect{j}}} \\
& = \frac{1}{2} \paren{2 \paren{\sum_{j \in V} w_j} \sum_{i\in V} w_i 
	\norm{\vect{i}}^2 
	- 2 \inprod{\sum_{i \in V} w_i \vect{i}, \sum_{i \in V} w_i \vect{i}}} \\
& = \paren{\paren{\sum_{j \in V} w_j} \sum_{i\in V} w_i \norm{\vect{i}}^2 
	- \inprod{\sum_{i \in V} w_i \vect{i}, \sum_{i \in V} w_i \vect{i}}} . 
\end{align*}
If $(\vect{i})_{i \in V_G \cup E_G \cup \set{\emptyset}}$ is a feasible solution
to SDP \ref{sdp:spar}, then it is easy to see that 
$(\vect{i} -c )_{i \in V_G \cup E_G \cup \set{\emptyset}}$ 
is also a feasible solution of the same cost for any vector $c$ (of the same dimension).
Fix $c := \paren{\sum_{i \in V} w_i \vect{i}}/\paren{\sum_{j \in V} w_j}$.
Then we can ensure that we get that $\sum_{i \in V} w_i \vect{i} = 0$. Moreover, from the calculation
above, we get that constraints \ref{eq:normalization} in SDP \ref{sdp:spar} reduces to 
$\sum_{i\in V} w_i \norm{\vect{i}}^2 = 1/\paren{\sum_{j \in V} w_j}$.
Constraints \ref{eq:c3-u} and \ref{eq:normalization-u} in SDP \ref{sdp:u} are 
these two constraints with $w_i = \deg_i \ \forall i \in V$ and with the different
normalization in the latter constraint (the different normalization in this 
constraint is the because of the different normalizations in the definition of 
sparsest cut (Definition \ref{def:sparsity}) and conductance 
(Definition \ref{def:conductance})); we write these constraints in this form 
since it will be more useful for our analysis in this form.
The $\ell_2^2$ triangle inequality constraints in SDP \ref{sdp:spar}
are excluded from SDP \ref{sdp:u} since they will not be used. 

We describe an algorithm to round the solution to the SDP relaxation.
It is based on computing an embedding of the vectors from the 
SDP solution on a random Guassian vector and using the resulting line embedding.

\begin{algorithm}
\caption{Randomized Rounding}
\label{alg:sdp-random-projection}
\begin{enumerate}
	\item Solve SDP \ref{sdp:spar} to obtain vectors 
		$(\vect{i})_{i \in V \cup E} \subset \R^{V \cup E}$.
	\item \label{step:random-projection}
	Sample $g \sim \cN\paren{0,1}^{V \cup E}$ and set $x_i \defeq \inprod{\vect{i},g}$.
	\item Compute a set $S$ from $x$ using the algorithm from Proposition \ref{prop:cheeger-u}.
	\item Output $S$.
\end{enumerate}
\end{algorithm}

\subsection{Bounding Directed Conductance using the Directed Rayleigh quotient}
Even though Theorem \ref{thm:alg1} is for symmetric conductance, we prove
some of our bounds more generally for directed conductance.
We define the notion of the {\em Rayleigh quotient} for polymatroidal networks
generalizing the notion of Rayleigh quotient for graphs (folklore), hypergraphs 
\cite{ChanLTZ18}, etc. 
See Section \ref{sec:related} for a brief discussion of other notions of conductance
and Rayleigh quotient.
\begin{definition}
Let $G = \paren{V,E,(w_i)_{i \in V},(\rhoplus{v}{\cdot},\rhominus{v}{\cdot})_{v \in V}}$
be polymatroidal network.
For a vector $x \in \R^{V \cup E}$, we define 
\[ \ral{G}{+}{x} \defeq \frac{\sum_{v \in V} \paren{\hrhoplus{v}{\dvright{+}{\ell_2^2}{x}{v}} 
	+ \hrhominus{v}{\dvleft{-}{\ell_2^2}{x}{v}}}}{\sum_{v \in V} \deg_v x_v^2 } \] 
and
\[ \ral{G}{-}{x} \defeq \frac{\sum_{v \in V} \paren{\hrhoplus{v}{\dvleft{+}{\ell_2^2}{x}{v}} 
	+ \hrhominus{v}{\dvright{-}{\ell_2^2}{x}{v}}}}{\sum_{v \in V} \deg_v x_v^2 } . \]
We define the {\em Rayleigh quotient} of $x$ as $\ral{G}{}{x} \defeq 
	\min \set{\ral{G}{+}{x},\ral{G}{-}{x}}$.
\end{definition}
The main technical proposition we prove is the following. 
\begin{proposition}
\label{prop:cheeger}
Let $G = \paren{V,E,(w_i)_{i \in V},(\rhoplus{v}{\cdot},\rhominus{v}{\cdot})_{v \in V}}$
be a polymatroidal network,
and let $x \in \R^{V \cup E}$ be any vector satisfying 
$\inprod{x_V, D \allones_V} = 0$. There exists a deterministic polynomial
time algorithm to compute a set $S$ satisfying 
	\[ \expn{}{G}{S} \leq 32 \sqrt{\ral{G}{}{x}} . \]
\end{proposition}
We have not optimized the constant in Proposition
\ref{prop:cheeger} and it could potentially be improved.
This proposition is similar to Lemma H.2 of \cite{LiM18}; it proves an 
analogous statement for submodular hypergraphs whereas Proposition \ref{prop:cheeger} proves 
this for directed polymatroidal networks.

We first prove some lemmas that we will use to prove Proposition \ref{prop:cheeger}.
We first observe the following.
\begin{lemma}
\label{lem:+-}
For any vector $x \in \R^{V \cup E}$, any vertex $v\in V$ and any norm $\ell$, we have
\[ \dvright{+}{\ell}{x}{v} = \dvleft{+}{\ell}{-x}{v} \qquad \textrm {and} \qquad
\dvright{-}{\ell}{x}{v} = \dvleft{-}{\ell}{-x}{v} 
\]
\end{lemma}

\begin{proof}
Fix an edge $(v,u) \in E$. Then 
\[ \ell \paren{\max \set{0, x_v - x_{(v,u)}}} = 
	\ell \paren{\max \set{0, (- x_{(v,u)}) - (-x_v)}} . \]
Therefore, $\dvright{+}{\ell}{x}{v} = \dvleft{+}{\ell}{-x}{v}$. 
The second equality is proved similarly.
\end{proof}
\begin{corollary}[Corollary to Lemma \ref{lem:+-}]
\label{cor:+-}
$\ral{G}{+}{x} = \ral{G}{-}{-x}$.
\end{corollary}

One of the steps in the proof of the Cheeger's inequality for graphs is to 
restrict the size of the support of the line-embedding vector; this is used to 
bound the volume of the set output by the algorithm (recall that we want to
output a set of volume at most half the volume of the vertex set). 
This is done by taking either the positive part or the negative part of
the vector and showing that Rayleigh quotient for at least one them can 
be bounded using the Rayleigh quotient of the whole vector.
We will need to
prove an analogous statement for polymatroidal networks which will require
a little more work. 

For a vector $x \in \R^{V \cup E}$, we define
\[ x^+_i \defeq \max\set{x_i,0} \qquad \textrm{and} \qquad x^-_i \defeq \max\set{0,-x_i} .  \]
\begin{lemma}
\label{lem:+-supp}
Let $v \in V$ be an arbitrary vertex and $x \in \R^{V \cup E}$ be an arbitrary vector.
Then the following hold $\forall a \in \set{x^+,-x^-},\ b \in \set{+,-}$.
\begin{enumerate}
\item $\hat{\rho}_v^b\paren{\dvright{b}{\ell_2^2}{a}{v}}
	\leq \hat{\rho}_v^b\paren{\dvright{b}{\ell_2^2}{x}{v}}$.
\item $\hat{\rho}_v^b\paren{\dvleft{b}{\ell_2^2}{a}{1}}
	\leq \hat{\rho}_v^b\paren{\dvleft{}{\ell_2^2}{x}{v}}$.
\end{enumerate}
\end{lemma}

\begin{proof}

\begin{claim}
\label{claim:+-}
For any $a,b \in \R$, we have 
\[ \max \set{0,a^+ - b^+} \leq \max \set{0, a - b}
\qquad \textrm{and} \qquad 
\max \set{0,\paren{-a^-} - \paren{-b^-}} \leq \max \set{0, a - b} . \]
\end{claim}
	
\begin{proof}[Proof of Claim \ref{claim:+-}]
\begin{enumerate}
\item If $a, b \geq 0$, then $a^+ - b^+ = a - b$. 
\item If $a, b < 0$, then $\max \set{0,a^+ - b^+} = 0 \leq \max \set{0, a - b}$.
\item If $a \geq 0,\ b < 0$, then $\max \set{0,a^+ - b^+} = a < a - b = \max \set{0, a - b}$
\item If $a < 0,\ b \geq 0$, then $\max \set{0,a^+ - b^+} = 0 = \max \set{0, a - b}$.
\end{enumerate}
Therefore, we have $\max \set{0,a^+ - b^+} \leq \max \set{0, a - b}$ in all the cases.
Similarly, one can prove that $\max \set{0,\paren{-a^-} - \paren{-b^-}} \leq \max \set{0, a - b}$.
\end{proof}

Using Claim \ref{claim:+-} coordinate-wise, we get that
$\forall a \in \set{x^+,-x^-},\ b \in \set{+,-}$,
\[ \dvright{b}{\ell_2^2}{a}{v} \leq \dvright{b}{\ell_2^2}{x}{v} 
	\qquad \textrm{and} \qquad 
\dvleft{b}{\ell_2^2}{a}{v} \leq \dvleft{b}{\ell_2^2}{x}{v} . \]
Using Fact \ref{fact:monotone-sub} finishes the proof of the lemma.
\end{proof}

\begin{lemma}
\label{lem:shift}
Given any vector $x \in \R^{V \cup E}$ satisfying $\inprod{x_V, D \allones_V} =$, 
there is a polynomial time algorithm to
compute a $y \in \R^{V \cup E}$ satisfying 
$\vol{}{\supp{y_V^+}},\vol{}{\supp{y_V^-}} \leq \vol{}{V}/2$
and
\begin{enumerate}
\item $ \min \set{\ral{}{+}{y^+}, \ral{}{+}{-y^-}} \leq 2 \ral{}{+}{x}$.
\item $\min \set{\ral{}{-}{y^+}, \ral{}{-}{-y^-}} \leq 2 \ral{}{-}{x}$. 
\item $ \min \set{\ral{}{+}{y^+} + \ral{}{-}{y^+}, \ral{}{+}{-y^-} + \ral{}{-}{-y^-}} 
	\leq 2 \paren{\ral{}{+}{x} + \ral{}{-}{x}}$.
\end{enumerate}
\end{lemma}

\begin{proof}
Let $y = x + c\allones$ for a constant $c$ such that 
$\vol{}{\supp{y_V^+}},\vol{}{\supp{y_V^-}} \leq \vol{}{V}/2$.
Then 
\[ \hrhoplus{v}{\dvright{+}{\ell_2^2}{x}{v}} 
+ \hrhominus{v}{\dvleft{-}{\ell_2^2}{x}{v}}
= \hrhoplus{v}{\dvright{+}{\ell_2^2}{y}{v}} 
+ \hrhominus{v}{\dvleft{-}{\ell_2^2}{y}{v}} 
\qquad \forall v \in V, \]
and using $0 = \inprod{x_V, D \allones_V} = \sum_{v \in V} \deg_v x_v$,
\[ \sum_{v \in V} \deg_v y_v^2 = \sum_{v \in V} \deg_v (x_v + c)^2 
\geq \sum_{v \in V} \deg_v x_v^2 + c^2 \sum_{v \in V} \deg_v + 2c \sum_{v \in V} \deg_v x_v 
	\geq \sum_{v \in V} \deg_v x_v^2 . \]
Therefore, using Lemma \ref{lem:+-supp} 
\begin{align*}
\ral{}{+}{x} & = \frac{\sum_{v \in V} \paren{\hrhoplus{v}{\dvright{+}{\ell_2^2}{x}{v}} 
	+ \hrhominus{v}{\dvleft{-}{\ell_2^2}{x}{v}}}}{\sum_{v \in V} \deg_v x_v^2} 
	 \geq \frac{\sum_{v \in V} \paren{\hrhoplus{v}{\dvright{+}{\ell_2^2}{y}{v}} 
	+ \hrhominus{v}{\dvleft{-}{\ell_2^2}{y}{v}}}}{\sum_{v \in V} \deg_v y_v^2} \\
& \geq \frac{1}{2} \cdot \frac{\sum_{v \in V} \paren{\hrhoplus{v}{\dvright{+}{\ell_2^2}{y^+}{v}} 
	+ \hrhominus{v}{\dvleft{-}{\ell_2^2}{y^+}{v}}} + 
	\sum_{v \in V} \paren{\hrhoplus{v}{\dvright{+}{\ell_2^2}{-y^-}{v}} 
	+ \hrhominus{v}{\dvleft{-}{\ell_2^2}{-y^-}{v}}}}
	{\sum_{v \in V} \deg_v \paren{y_v^+}^2 + \sum_{v \in V} \deg_v \paren{-y_v^-}^2} \\
& \geq \frac{1}{2} \min \set{
	\frac{\sum_{v \in V}\paren{\hrhoplus{v}{\dvright{+}{\ell_2^2}{y^+}{v}} 
	+ \hrhominus{v}{\dvleft{-}{\ell_2^2}{y^+}{v}}}}
	{\sum_{v \in V} \deg_v \paren{y_v^+}^2},
	\frac{\sum_{v \in V}\paren{\hrhoplus{v}{\dvright{+}{\ell_2^2}{-y^-}{v}} 
	+ \hrhominus{v}{\dvleft{-}{\ell_2^2}{-y^-}{v}}}}
	{\sum_{v \in V} \deg_v \paren{-y_v^-}^2}} \\
& = \frac{1}{2} \min \set{\ral{}{+}{y^+}, \ral{}{+}{-y^-}} .
\end{align*}
The other parts are proved similarly.
\end{proof}

\begin{lemma}
\label{lem:cs}
For any $z \in \Rgeq^{V \cup E}$,
\[ \frac{\sum_{v \in V} \paren{ \hrhoplus{v}{\dvright{+}{\ell_1}{z^2}{v}} 
	+ \hrhominus{v}{\dvleft{-}{\ell_1}{z^2}{v}}}}
{\sum_{v \in V} \deg_v z_v^2}
	\leq 16 \sqrt{\ral{}{+}{z}} \]
and 
\[ \frac{\sum_{v \in V}\paren{\hrhoplus{v}{\dvleft{+}{\ell_1}{z^2}{v}} 
	+ \hrhominus{v}{\dvright{-}{\ell_1}{z^2}{v}}}}
{\sum_{v \in V} \deg_v z_v^2}
	\leq 16 \sqrt{\ral{}{-}{z}} . \]
\end{lemma}

\begin{proof}
For $v \in V$, let $\delta^+(v) = \set{(v,1), \ldots, (v,r_v)}$ such that
$z_{(v,1)} \leq \cdots \leq z_{(v,r_v)}$. 
Let $r_v'$ be the index such that 
$z_{(v,1)} \leq \cdots \leq z_{(v,r_v')} \leq z_v 
< z_{(v,r_v' + 1)} \leq \cdots \leq z_{(v,r_v)}$.
	Then, 
$z_{(v,1)}^2 \leq \cdots \leq z_{(v,r_v')}^2 \leq z_v^2 
< z_{(v,r_v' + 1)}^2 \leq \cdots \leq z_{(v,r_v)}^2$
and,
\begin{equation}
\label{eq:sign+}
z_v^2 - z_{(v,1)}^2 \geq \cdots \geq  z_v^2 - z_{(v,r_v')}^2 \geq 0
>  z_v^2 - z_{(v,r_v'+1)}^2 \geq  z_v^2 - z_{(v,r_v)}^2 . 
\end{equation}

\begin{flalign}
& \sum_{v \in V}\hrhoplus{v}{\dvright{+}{\ell_1}{z^2}{v}} & \nonumber \\
& = \sum_{v \in V} \left( \sum_{i \in [r_v']}\paren{ 
	\rhoplus{v}{\set{(v,1),\ldots,(v,i)}} - \rhoplus{v}{\set{(v,1),\ldots,(v,i-1)}}
	\paren{z_v^2 - z_{(v,i)}^2}} \right) \nonumber \\
& = \sum_{v \in V} \sum_{i \in [r_v']}\paren{ 
	\rhoplus{v}{\set{(v,1),\ldots,(v,i)}} - \rhoplus{v}{\set{(v,1),\ldots,(v,i-1)}}
	\paren{z_v - z_{(v,i)}} \paren{z_v + z_{(v,i)}} } \label{eq:dvr0}
\end{flalign}
Since $\rhoplus{v}{\cdot}$ is a monotone function, we get that 
\begin{equation*}
\rhoplus{v}{\set{(v,1),\ldots,(v,i)}} - \rhoplus{v}{\set{(v,1),\ldots,(v,i-1)}} \geq 0
\qquad \forall i \in [r_v],\ v \in V 
\end{equation*}
From equation \ref{eq:sign+}, we that $\paren{z_v - z_{(v,i)}} \geq 0
\ \forall i \in [r_v']$, and $\paren{z_v + z_{(v,i)}} \geq 0$.
Therefore, using these and the Cauchy-Schwarz inequality in \ref{eq:dvr0}
\begin{flalign}
& \sum_{v \in V}\hrhoplus{v}{\dvright{+}{\ell_1}{z^2}{v}} \nonumber \\ 
& \leq \left( \sum_{v \in V} \left( \sum_{i \in [r_v']}\paren{ 
	\rhoplus{v}{\set{(v,1),\ldots,(v,i)}} - \rhoplus{v}{\set{(v,1),\ldots,(v,i-1)}}
	\paren{z_v - z_{(v,i)}}^2 } \right) \right)^{1/2} \nonumber \\
& \hspace{1cm} \cdot \left( \sum_{v \in V} \left( \sum_{i \in [r_v']}\paren{ 
	\rhoplus{v}{\set{(v,1),\ldots,(v,i)}} - \rhoplus{v}{\set{(v,1),\ldots,(v,i-1)}}
	\paren{z_v + z_{(v,i)}}^2 } \right) \right)^{1/2} \label{eq:dvr1} 
\end{flalign}
Now, the first term of equation \ref{eq:dvr1} gives
\begin{flalign}
& \left( \sum_{v \in V} \left( \sum_{i \in [r_v']}\paren{ 
	\rhoplus{v}{\set{(v,1),\ldots,(v,i)}} - \rhoplus{v}{\set{(v,1),\ldots,(v,i-1)}}
	\paren{z_v - z_{(v,i)}}^2 } \right) \right)^{1/2} \nonumber \\ 
& = \sqrt{\sum_{v \in V} \hrhoplus{v}{\dvright{+}{\ell_2^2}{z}{v}}} 
\label{eq:dvr2}
\end{flalign}
The second term of equation \ref{eq:dvr1} gives
\begin{flalign}
& \left( \sum_{v \in V} \left( \sum_{i \in [r_v']}\paren{ 
	\rhoplus{v}{\set{(v,1),\ldots,(v,i)}} - \rhoplus{v}{\set{(v,1),\ldots,(v,i-1)}}
	\paren{\paren{z_{(v,i)} - z_v} + 2 z_v}^2 } \right) \right)^{1/2} \nonumber \\
& \leq \left(2 \sum_{v \in V} \left( \sum_{i \in [r_v']}\paren{ 
	\rhoplus{v}{\set{(v,1),\ldots,(v,i)}} - \rhoplus{v}{\set{(v,1),\ldots,(v,i-1)}}
	\paren{\paren{z_{(v,i)} - z_v}^2 + 4 z_v^2}}  \right) \right)^{1/2} \nonumber \\
& = \left(2 \sum_{v \in V} \hrhoplus{v}{\dvright{+}{\ell_2^2}{z}{v}}
	+ 8 \sum_{v \in V} z_v^2 \left( \sum_{i \in [r_v']}\paren{ 
	\rhoplus{v}{\set{(v,1),\ldots,(v,i)}} - \rhoplus{v}{\set{(v,1),\ldots,(v,i-1)}}
	} \right) \right)^{1/2} \nonumber \\
& \leq \sqrt{2 \sum_{v \in V} \hrhoplus{v}{\dvright{+}{\ell_2^2}{z}{v}}
	+ 8 \sum_{v \in V} \rhoplus{v}{\set{(v,1),\ldots,(v,r_v)}} z_v^2} 
	\nonumber \\  
& \leq \sqrt{2 \sum_{v \in V} \hrhoplus{v}{\dvright{+}{\ell_2^2}{z}{v}}}
	+ \sqrt{8 \sum_{v \in V} \deg_v z_v^2} \label{eq:dvr3} 
\end{flalign}
Therefore, from equations \ref{eq:dvr2} and \ref{eq:dvr3}, we get 
\begin{equation}
\label{eq:dvr4}
\sum_{v \in V}\hrhoplus{v}{\dvright{+}{\ell_1}{z^2}{v}} 
\leq 2 \sum_{v \in V} \hrhoplus{v}{\dvright{+}{\ell_2^2}{z}{v}} +
	4 \sqrt{\sum_{v \in V} \hrhoplus{v}{\dvright{+}{\ell_2^2}{z}{v}}}
	\sqrt{\sum_{v \in V} \deg_v z_v^2} 
\end{equation}
Similarly, we get
\begin{equation}
\label{eq:dvr5}
\sum_{v \in V}\hrhominus{v}{\dvleft{-}{\ell_1}{z^2}{v}} 
\leq 2 \sum_{v \in V} \hrhominus{v}{\dvleft{-}{\ell_2^2}{z}{v}} +
	4 \sqrt{\sum_{v \in V} \hrhominus{v}{\dvleft{-}{\ell_2^2}{z}{v}}}
	\sqrt{\sum_{v \in V} \deg_v z_v^2} 
\end{equation}
Therefore, using equations \ref{eq:dvr4} and \ref{eq:dvr5}
\begin{flalign}
& \frac{\sum_{v \in V}\hrhoplus{v}{\dvright{+}{\ell_1}{z^2}{v}} 
	+ \sum_{v \in V}\hrhominus{v}{\dvleft{-}{\ell_1}{z^2}{v}}}
	{\sum_{v \in V} \deg_v z_v^2} & \nonumber \\
	& \leq 4 \frac{
	\sum_{v \in V} \paren{\hrhoplus{v}{\dvright{+}{\ell_2^2}{z}{v}} +
	\sum_{v \in V} \hrhominus{v}{\dvleft{-}{\ell_2^2}{z}{v}}}}
	{\sum_{v \in V} \deg_v z_v^2} & \nonumber \\
	& \hspace{1 cm} + 4\frac{\paren{\sqrt{\sum_{v \in V}\hrhoplus{v}{\dvright{+}{\ell_2^2}{z}{v}}} 
	+ \sqrt{\sum_{v \in V}\hrhominus{v}{\dvleft{-}{\ell_2^2}{z}{v}}}}
	\sqrt{\sum_{v \in V} \deg_v z_v^2}}{ \sum_{v \in V} \deg_v z_v^2} 
	\nonumber \\
	& \leq 8 \paren{\ral{}{+}{z} + \sqrt{\frac{\sum_{v \in V} \paren{\hrhoplus{v}{\dvright{+}{\ell_2^2}{z}{v}}
	+ \sum_{v \in V}\hrhominus{v}{\dvleft{-}{\ell_2^2}{z}{v}}}}
	{\sum_{v \in V} \deg_v z_v^2}}} \nonumber \\
	& \qquad \qquad \textrm{(using $\sqrt{a} + \sqrt{b} \leq 2 \sqrt{a + b}\ \forall a,b \geq 0$
	 )}& \nonumber \\
	 & \leq 8 \paren{\ral{}{+}{z} + \sqrt{\ral{}{+}{z}}} . \nonumber 
\end{flalign}
The second part is proved similarly.
\end{proof}

\begin{proof}[Proof of Proposition \ref{prop:cheeger}]
Let $y$ be the vector obtained from $x$ using Lemma \ref{lem:shift}; we have that
\begin{equation}
\label{eq:r+-}
	\min \set{\ral{}{+}{y^+}, \ral{}{+}{-y^-}} \leq 2 \ral{}{+}{x} . 
\end{equation}
Let us first consider the case when $y^+$ is the minimizer of the quantity above.
Let $z = y^+$.
Using Lemma \ref{lem:cs} we get that 
\[ \frac{\sum_{v \in V}\paren{\hrhoplus{v}{\dvright{+}{\ell_1}{z^2}{v}} 
	+ \hrhominus{v}{\dvleft{-}{\ell_1}{z^2}{v}}}}
{\sum_{v \in V} \deg_v z_v^2}
	\leq 32 \sqrt{\ral{}{+}{x}} . \]
Next, let us consider the case when $-y^-$ is the minimizer of the 
quantity in equation \ref{eq:r+-}. Let $z = y^-$; note that $z \geq 0$. 
Using Lemma \ref{lem:+-} we get
that 
\begin{align*}
\ral{}{-}{y^-} & =	
\frac{\sum_{v \in V}\paren{ \hrhoplus{v}{\dvleft{+}{\ell_2^2}{y^-}{v}} 
	+ \hrhominus{v}{\dvright{-}{\ell_2^2}{y^-}{v}}}}
	{2 \sum_{v \in V} \deg_v \paren{y_v^-}^2}  
 = \frac{\sum_{v \in V} \paren{\hrhoplus{v}{\dvright{+}{\ell_2^2}{-y^-}{v}} 
	+ \hrhominus{v}{\dvleft{-}{\ell_2^2}{-y^-}{v}}}}
	{2 \sum_{v \in V} \deg_v \paren{-y_v^-}^2} \\
& \leq 2 \ral{}{+}{x} .
\end{align*}
Using Lemma \ref{lem:cs} we get that 
\begin{equation}
\frac{\sum_{v \in V}\paren{\hrhoplus{v}{\dvleft{+}{\ell_1}{z^2}{v}} 
	+ \hrhominus{v}{\dvright{-}{\ell_1}{z^2}{v}}}}
	{\sum_{v \in V} \deg_v z_v^2}
	\leq 16 \sqrt{\ral{}{-}{z}} = 16 \sqrt{\ral{}{-}{y^-}} 
	\leq 32 \sqrt{\ral{}{+}{x}} .
\end{equation}
Therefore, using the algorithm from Lemma \ref{lem:line-embedding} on the vector $z^2$
finishes the proof of this proposition in both cases. 
\end{proof}

\subsection{Bounding Symmetric Conductance using the Symmetric Rayleigh quotient}
We define the ``symmetric'' Rayleigh quotient of a vector $x \in \R^{V \cup E}$ 
as $\ralu{G}{x} \defeq \ral{G}{+}{x} + \ral{G}{-}{x}$.

\begin{proposition}
\label{prop:cheeger-u}
Let $G = \paren{V,E,(w_i)_{i \in V},(\rhoplus{v}{\cdot},\rhominus{v}{\cdot})_{v \in V}}$
be a polymatroidal network 
and let $x \in \R^{V \cup E}$ be a vector satisfying 
$\inprod{x_V, D \allones_V} = 0$. There exists a deterministic polynomial
time algorithm to compute a set $S$ satisfying 
	\[ \uexpn{G}{S} \leq 128 \sqrt{\ralu{G}{x}} .\] 
\end{proposition}

\begin{proof}
Let $y$ be the vector obtained from $x$ using item 3 of Lemma \ref{lem:shift}; 
we have that
\[ \min \set{\ral{}{+}{y^+} + \ral{}{-}{y^+}, \ral{}{+}{-y^-} + \ral{}{-}{-y^-}} 
	\leq 2 \ral{}{+}{x} . \]
Using Corollary \ref{cor:+-} we get that 
\[ \min \set{\ral{}{+}{y^+} + \ral{}{-}{y^+}, \ral{}{+}{y^-} + \ral{}{-}{y^-}} 
	\leq 2 \ral{}{+}{x} . \]
Let $z \in \set{y^+, y^-}$ be the minimizer of the quantity above.
\begin{flalign}
& \frac{\sum_{v \in V}\paren{\hrhoplus{v}{\dv{+}{\ell_1}{z^2}{v}} 
	+ \hrhominus{v}{\dv{-}{\ell_1}{z^2}{v}}}}
	{\sum_{v \in V} \deg_v z_v^2} \nonumber \\ 
& \leq \frac{\sum_{v \in V}\paren{\hrhoplus{v}{\dvright{+}{\ell_1}{z^2}{v}} 
	+ \hrhoplus{v}{\dvleft{+}{\ell_1}{z^2}{v}} 
	+ \hrhominus{v}{\dvright{-}{\ell_1}{z^2}{v}}
	+ \hrhominus{v}{\dvleft{-}{\ell_1}{z^2}{v}}}}
	{\sum_{v \in V} \deg_v z_v^2} \quad 
	\paren{\textrm{using Lemma \ref{lem:summ}}} \nonumber \\ 
& \leq 32 \paren{\sqrt{\ral{}{+}{x}} + \sqrt{\ral{}{-}{x}}} \quad 
	\paren{\textrm{using Lemma \ref{lem:cs}}} \nonumber \\ 
	& \leq 64 \paren{\sqrt{\ralu{}{x}}}  \qquad \paren{\textrm{using }
	\sqrt{a} + \sqrt{b} \leq 2 \sqrt{a + b} \ \forall a,b \in Rgeq} . 
\end{flalign}
Therefore, using the algorithm from Lemma \ref{lem:line-embedding-u} on the vector $z^2$
finishes the proof of this proposition in both cases. 

\end{proof}

\subsection{Spectral Approximation Algorithm}
We need to bound the Rayleigh quotient of $x$ (from Step \ref{step:random-projection}
of Algorithm \ref{alg:sdp-random-projection}). Towards this, we will use the following lemma.
\begin{lemma}
\label{lem:logd-approx}
Let $a_1, \ldots, a_d \in \R^n$ be an arbitrary set of vectors and $\basepoly{\rho} \subset \R^d$
be the base polytope of a monotone submodular function $\rho$. Then
\[ \Ex{g \sim \cN(0,1)^n}{\hrho{}{\inprod{a_1, g}^2, \ldots, \inprod{a_d, g}^2}}
= \bigo{\log d} \cdot \hrho{}{\norm{a_1}^2, \ldots, \norm{a_d}^2 } .\]
\end{lemma}

\begin{proof}
Since $\basepoly{\rho}$ is the base polytope of a monotone submodular function, we know that 
the optimal $w$ has only non-negative values. 
\begin{flalign}
& \Ex{g \sim \cN(0,1)^n}{\hrho{}{\inprod{a_1, g}^2, \ldots, \inprod{a_d, g}^2}} & \nonumber \\
	& = \Ex{g \sim \cN(0,1)^n}{\sup_{\substack{w \in \basepoly{\rho} \\ w \geq 0}}  
\inprod{w,\sqbracs{\inprod{a_1, g}^2, \ldots, \inprod{a_d, g}^2 }^T} } &
	\textrm{(using Lemma \ref{lem:lov-ext})} \nonumber \\
& = \Ex{g \sim \cN(0,1)^n}{\sup_{\substack{w \in \basepoly{\rho} \\ w \geq 0}}
	\paren{\sum_{i \in [d]} w_i \norm{a_i}^2 \inprod{\frac{a_i}{\norm{a_i}}, g}^2}} 
	& \nonumber \\
& \leq \Ex{g \sim \cN(0,1)^n}{\sup_{\substack{w \in \basepoly{\rho} \\ w \geq 0}}
		\paren{\sum_{i \in [d]} 
	w_i \norm{a_i}^2 } \max_{i \in [d]} \inprod{\frac{a_i}{\norm{a_i}}, g}^2} 
	& \textrm{(using $w_i \geq 0 \ \forall i \in [d]$)}
	\nonumber \\
& = \sup_{w \in \basepoly{\rho}} \paren{\sum_{i \in [d]} w_i \norm{a_i}^2 } \Ex{g \sim \cN(0,1)^n}{\max_{i \in [d]} \inprod{\frac{a_i}{\norm{a_i}}, g}^2} & \nonumber \\
	& = \bigo{\log d} \cdot \hrho{}{\norm{a_1}^2, \ldots, \norm{a_d}^2 } 
	& \textrm{(using Fact \ref{fact:maxgauss})} \nonumber
\end{flalign}
\end{proof}

\begin{fact}[Folklore]
\label{fact:maxgauss}
Let $g_1, \ldots, g_d$ be (not necessarily independent) Gaussian random variables
satisfying $\Ex{}{g_i} = 0 \ \forall i \in [d]$. Then 
\[ \Ex{}{\max_{i \in [d]} g_i^2} = \bigo{\log d} \cdot \max_{i \in [d]} \Ex{}{g_i^2} .\]
\end{fact}

\begin{proof}[Proof of Theorem \ref{thm:alg1}]
Using Lemma \ref{lem:logd-approx} we get
\begin{equation} 	
\label{eq:numerator}
	\Ex{g}{\sum_{v \in V} \paren{\hrhoplus{v}{\dv{+}{\ell_2^2}{x}{v}} 
	+ \hrhominus{v}{\dv{-}{\ell_2^2}{x}{v}}}}
	= \bigo{\log r} \cdot \sum_{v \in V} \paren{\hrhoplus{v}{\dv{+}{\ell_2^2}{\vect{}}{v}} 
	+ \hrhominus{v}{\dv{-}{\ell_2^2}{\vect{}}{v}}}
	= \bigo{\log r} \cdot \sdpval .	
\end{equation}
Next, 
\[ \Ex{g}{\sum_{i \in V} \deg_i x_i^2} 
	= \sum_{i \in V} \deg_i \vect{i}^T \Ex{g}{gg^T} \vect{i}
	=  \sum_{i \in V} \deg_i \norm{\vect{i}}^2 . \] 
Using the Paley-Zygmund inequality, it is well known that 
(for e.g., see Lemma 9.8 of \cite{LouisRV13})
\begin{equation}
\label{eq:denominator}
\Prob{g}{\sum_{i \in V} \deg_i x_i^2 \geq 
	\frac{1}{2} \paren{\sum_{i \in V} \deg_i \norm{\vect{i}}^2}} =
\Prob{g}{\sum_{i \in V} \deg_i x_i^2 \geq 
	\frac{1}{2} \Ex{g}{\sum_{i \in V} \deg_i x_i^2 }} \geq \frac{1}{12} . 
\end{equation}
Therefore, using Markov's inequality with equation \ref{eq:numerator}, 
equation \ref{eq:denominator} and the union bound, we get that 
$\ralu{G}{x} = \bigo{\log r} \cdot \sdpval$ with constant probability.
Moreover, using constraint \ref{eq:c3-u} in SDP \ref{sdp:u},
\[ \sum_{i \in V} \deg_i x_i = \inprod{\sum_{i \in V} \deg_i \vect{i},g} = 0 .  \]
Therefore, using Proposition \ref{prop:cheeger-u}, we
get a set $S$ such that $\uexpn{G}{S} = \bigo{\sqrt{\sdpval \log r}}$ with constant
probability.
\end{proof}

\subsection{Undirected Polymatroidal Networks}

We show in Corollary \ref{cor:alg1} that Theorem \ref{thm:alg1} implies the 
same approximation guarantee for conductance in undirected polymatroidal 
networks as well.

\begin{corollary}[Corollary to Theorem \ref{thm:alg1}]
\label{cor:alg1}
Let $G = \paren{V,E,(w_i)_{i \in V},(\rho_{v})_{v \in V}}$
be an undirected polymatroidal network, and let
$r \defeq \max_{v \in V} \Abs{\delta(v)}$ be the maximum degree in $G$.
There is a randomized polynomial time algorithm that outputs a set $S \subset V$ 
such that $\expn{}{G}{S} = \bigo{\sqrt{\expnG{}{G} \log r}}$ with constant
probability.
\end{corollary}

\begin{proof}
We construct a directed polymatroidal network
$G' = \paren{V,E',(w_i)_{i \in V},(\rhoplussymb{v},\rhominussymb{v})_{v\in V}}$
as follows. 
For each $\set{u,v} \in E$, add edges $(u,v),(v,u)$ to $E'$.
For a vertex $v \in V$, we define $(\rhoplussymb{v},\rhominussymb{v})$
as follows. For $S \subseteq \delta^+_{G'}(v)$, define 
$\rhoplus{v}{S} \defeq \rho_v(S')$ where $S'$ denotes the set of
corresponding undirected edges in $G$. 
For $S \subseteq \delta^-_{G'}(v)$, define 
$\rhominus{v}{S} \defeq \rho_v(S')$ where $S'$ denotes the set of
corresponding undirected edges in $G$.

Now, fix $S \subset V$ and $g: \delta^+_{G'}(S) \to V$.
We define $\bar{g} : \delta_G(S) \to V$ from $g$ in the natural way:
for an edge $\set{u,v} \in \delta_G(S)$ with $u \in V, \ v \in V \setminus S$,
define $\bar{g}\paren{\set{u,v}} = g \paren{\paren{u,v}}$.
Then
\begin{align*}
\cut{G'}{\delta^+_{G'}(S),g} & = 
\sum_{v \in V} \paren{\rhoplus{v}{\delta^+_{G'}(v) \cap g^{-1}(v)} 
	+ \rhominus{v}{\delta^-_{G'}(v) \cap g^{-1}(v)}} \\
& = \sum_{v \in S} \paren{\rhoplus{v}{\delta^+_{G'}(v) \cap g^{-1}(v)}}
+ \sum_{v \in V \setminus S} \paren{\rhominus{v}{\delta^-_{G'}(v) \cap g^{-1}(v)}} \\
& = \sum_{v \in V} \paren{\rho_v\paren{\delta_{G}(v) \cap \bar{g}^{-1}(v)}} \\
& = \cut{G}{\delta_{G}(S),\bar{g}} .
\end{align*}
Therefore, 
\[ \cut{G'}{\delta^+_{G'}(S)} = 
	\min_{g: \set{\delta^+_{G'}(S) \to V}} \cut{G'}{\delta^+_{G'}(S),g} 
= \min_{\bar{g}: \set{\delta_{G}(S) \to V}} \cut{G}{\delta_{G}(S),\bar{g}} 
= \cut{G}{\delta_{G}(S)} . \]
Similarly, we can prove that
$ \cut{G'}{\delta^-_{G'}(S)} = \cut{G}{\delta_{G}(S)}$, 
and hence 
\begin{equation}
\label{eq:num-u}
\cut{G'}{\delta^+_{G'}(S)} + \cut{G'}{\delta^-_{G'}(S)}
	= 2\ \cut{G}{\delta_{G}(S)} .
\end{equation}
Moreover,
\begin{equation}
\label{eq:denom-u}
\vol{G'}{S} = \sum_{i \in S} \paren{\rhoplus{v}{\delta^+(S)} 
	+ \rhominus{v}{\delta^-(S)}} 
	= 2 \sum_{i \in S} \rho_v \paren{\delta(S)} 
	= 2\ \vol{G}{S} . 
\end{equation}
Using \ref{eq:num-u} and \ref{eq:denom-u} we get that 
	\[ \expn{}{G'}{S} = \expn{}{G}{S} \qquad \forall S \subseteq V. \]
Therefore, using Theorem \ref{thm:alg1} finishes this proof.
\end{proof}

\section{Polymatroidal Hypergraphs}
\label{sec:h}
Given a hypergraph $H = (V,E)$ with non-negative vertex weights $(w_i)_{i \in V}$,
and monotone submodular cut functions $\fplus{e}{\cdot}, \fminus{e}{\cdot}: 
2^e \to \Rgeq$ for each $e \in E$.
Let $n = \Abs{V}$, $m = \Abs{E}$, $r$ be the cardinality of the largest
hyperedge in $E$. For a set $S \subseteq V$, define 
$\delta(S) \defeq \set{e \in E: \emptyset \neq e \cap S \neq e}$.
The degree of an edge $e \in E$ is defined as $\deg_e \defeq \fplus{e}{e} 
+ \fminus{e}{e}$.

\begin{definition}
\label{def:spar-h}
Fix a set $S \subset V$. We define $\cut{H}{S}$ as
\[ \cut{H}{S} \defeq \sum_{e \in \delta_H(S)} \min\set{\fminus{e}{e \cap S},
	\fplus{e}{e \cap \paren{V \setminus S}}} . \] 
We define the sparsity of $S$, denoted by $\spar(S)$, as 
\[ \spar_H(S) \defeq \frac{\cut{H}{S}}{w(S) \cdot w(V\setminus S)} . \] 
We define the sparsity of the hypergraph as 
$ \spar_H \defeq \min_{S: \emptyset \neq S \neq V} \spar_H(S)$.
\end{definition}

Analogous to our definition of symmetric conductance of polymatroidal networks,
we define the symmetric conductance of polymatroidal hypergraphs as follows.
For some technical reasons which we will explain soon, we define conductance 
for subsets of $V \cup E$ and not just subsets of $V$.
\begin{definition}
\label{def:expn-h}
Fix a set $S \subset V \cup E$ such that for each $e \in S$, $e \cap S \neq \emptyset$,
i.e., each edge in $S$ contains at least one vertex in $S$. 
We define the volume of $S$ as $\vol{H}{S} = \sum_{e \in S \cap E} \deg_e$.
We define $\symcut{H}{S}$ as
\[ \symcut{H}{S} \defeq \sum_{e \in S \cap \delta_H(S \cap V)} 
	\paren{\fminus{e}{e \cap \paren{V \setminus S}} 
	+ \fplus{e}{e \cap \paren{V \setminus S}}} . \]
We define the symmetric conductance of $S$, denoted by $\uexpn{H}{S}$, as
\[ \uexpn{H}{S} \defeq \frac{\symcut{H}{S} + \symcut{H}{(V \cup E) \setminus S}}{\vol{H}{S}} \] and
we define the symmetric conductance of the hypergraph as 
\[ \uexpnG{H} \defeq \min_{\substack{ S \subset V \cup E: \\ 
	0 < \vol{H}{S} \leq \vol{H}{V}/2}} \uexpn{H}{S}. \] 
\end{definition}

The definition of conductance depends on the volume of the set. Since
the cut functions $\set{\fplus{e}{\cdot}, \fminus{e}{\cdot}}_{e \in E}$ 
are defined over subsets of edges, the natural analog of the definition
of degree in polymatroidal networks is to define degree for 
hyperedges. Therefore, we define conductance for subsets of $V \cup E$
and not just for subsets of $V$. 

The folklore definition of degree of a vertex in a hypergraph is the number
of hyperedges the vertex belongs to. Based on this definition of degree,
one can define the volume for subsets of vertices and hence, the conductance
of subsets of vertices. 
Generalizing this, one can define the degree of a vertex $v$ in a polymatroidal
hypergraph as $\deg_v \defeq \sum_{\substack{e \in E: \\ e \ni v}} \deg_e$
and define conductance for subsets of vertices based on this. 
However, a major drawback of these definitions is that for a subset $S \subseteq V$,
a hyperedge $e$ ``contributes'' to the vertex-degrees of $\Abs{e \cap S}$ vertices 
in $S$ where $\Abs{e \cap S} \in \set{0, \ldots, r}$. 
\cite{LouisM16,Louis15,ChanLTZ18} compensated for this in various ways;
see Section \ref{sec:related} for a discussion of their results. 
Our definition of symmetric conductance can be viewed as another approach 
to avoid this over-counting.

We will use the following standard reduction from \cite{Lawler73} to construct a
polymatroidal network from a polymatroidal hypergraph; this is basically the bipartite representation of hypergraphs
converted into a bidirected graph.
\begin{definition}[Factor graphs \cite{Lawler73}]
\label{def:reduction}
For each vertex in $i \in V_H$, we add a vertex $\inG{i}$ with weight $w_i$ to
$V_G$, and for each edge $e \in E_H$, we add a vertex $\inG{e}$ with weight $0$ 
to $V_G$. 
For each pair of vertices $\inG{i}, \inG{e} \in V_G$ we add edges
$(\inG{i}, \inG{e})$ and $(\inG{e}, \inG{i})$ if $i \in_H e$. 
For each $e \in E_H$, we add a submodular capacity functions 
	$\rhoplus{\inG{e}}{\cdot}, \rhominus{\inG{e}}{\cdot}$ of $\inG{e}$ defined as
\[ \rhoplus{\inG{e}}{\set{\paren{\inG{e}, \inG{a_1}}, \ldots, \paren{\inG{e}, \inG{a_t}}}} 
	\defeq \fplus{e}{\set{a_1, \ldots, a_t}}, \] 
\[ \rhominus{\inG{e}}{\set{\paren{\inG{a_1},\inG{e}}, \ldots, \paren{\inG{a_t},\inG{e}}}} 
	\defeq \fminus{e}{\set{a_1, \ldots, a_t}}, \]
and for each $v \in V_H$, we add a submodular capacity functions 
	$\rhoplus{\inG{v}}{\cdot}, \rhominus{\inG{v}}{\cdot}$ of $\inG{v}$ defined as
$ \rhoplus{\inG{v}}{\cdot} \defeq  \infty$, $\rhominus{\inG{v}}{\cdot} \defeq \infty $. 
\end{definition}

For any $S \subseteq V_H \cup E_H$, we define $\inG{S} \defeq \set{\inG{i} : i \in S}$.

\begin{restatable}{lemma}{HtoG}(\cite{Lawler73})
\label{lem:H-to-G}
For any $\alpha, t \in \Rplus$, there exists a set $S\subset V_H$ such that 
$\cut{H}{S} \leq \alpha$  and $w_H(S) \ w_H(V_H \setminus S) = t (n-t)$ if and only if 
there exists $T \subset V_G$ such that 
$\min \set{\cut{G}{\delta^+(T)},\cut{G}{\delta^-(T)}} \leq \alpha$ and
$w_{G}(T) \ w_G (V_G \setminus T)= t (n - t)$.
\end{restatable}
We prove this lemma in Section \ref{sec:H-to-G-proofs}. As direct corollaries,
we get the following. 
\begin{corollary}
\label{cor:H-to-G}
$\spar_G = \spar_H$. 
\end{corollary}

\begin{restatable}{lemma}{HtoGexpn}(\cite{Lawler73})
\label{lem:H-to-G-expn}
For any $S \subset V_G$ satisfying $e \in S \implies e \cap S \neq \emptyset$,
and $T \defeq \set{i \in V_H \cup E_H : \inG{i} \in S}$,
we have 
\[ \symcut{H}{T} + \symcut{H}{(V_H \cup E_H) \setminus T} 
	= \cut{G}{\delta^+(S)} + \cut{G}{\delta^-\paren{S}}, \]
and $\vol{H}{T} = \vol{G}{S \cap \inG{E_H}}$. 
\end{restatable}

\begin{corollary}
\label{cor:H-to-G-expn}
$\uexpnG{G} \leq \uexpnG{H}$.
\end{corollary}

\subsection{Sparsest Cut in Polymatroidal Hypergraphs}

In this section, we prove Theorem \ref{thm:mult-approx-h}.
Our proof is based on Algorithm \ref{alg:multapprox-h}.

\multapproxh*

\begin{algorithm}
\caption{ARV rounding algorithm}
\label{alg:multapprox-h}
\begin{enumerate}
\item Construct a polymatroidal flow network $G$ from $H$ using Definition \ref{def:reduction}.
\item Compute a set $T \subset V_G$ using the algorithm from 
	Theorem \ref{thm:mult-approx}.
\item Compute $S \subset V_H$ from $T$ using the algorithm in Lemma \ref{lem:H-to-G}.
\item Output $S$.
\end{enumerate}
\end{algorithm}

\begin{proof}[Proof of Theorem \ref{thm:mult-approx-h}]
The proof of this theorem is based on Algorithm \ref{alg:multapprox-h}.
Using Corollary \ref{cor:H-to-G}, we get that $\spar_G \leq \spar_H$.
Since $w_{\inG{e}} = 0 \ \forall e \in E$, we get that
$\Abs{\set{i \in V_G: w_i > 0}} \leq \Abs{V_H} = n$.
Therefore, using Theorem \ref{thm:mult-approx} we get that $\spar_G\paren{T}
= \bigo{\sqrt{\log n}} \spar_G = \bigo{\sqrt{\log n}} \spar_H$.
Finally, using Lemma \ref{lem:H-to-G} again we get that
$\spar_H(S) \leq \spar_G\paren{T} = \bigo{\sqrt{\log n}} \spar_H$.
\end{proof}

\subsection{Symmetric Conductance in Polymatroidal Hypergraphs}

\talgcheegerh*

The proof of Theorem \ref{thm:alg1-h} uses Algorithm \ref{alg:sdp-random-projection}
for rounding SDP \ref{sdp:u-h}.

\begin{figure}[htb]
\centering
\begin{boxedminipage}[htb]{0.8\linewidth}
\begin{sdp}
\label{sdp:u-h}
	\[ \min \sum_{v \in V_G} \paren{\hrhoplus{v}{\dv{+}{\ell_2^2}{\vect}{v}} 
	+ \hrhominus{v}{\dv{-}{\ell_2^2}{\vect}{v}}}. \]
	subject to:
\begin{align}
	\sum_{i \in E_H} \deg_{\inG{i}} \vect{\inG{i}} & = 0 \label{eq:c3-u-h} \\
	\sum_{i \in E_H} \deg_{\inG{i}} \norm{\vect{\inG{i}}}^2 & = 1 \label{eq:normalization-u-h}
\end{align}
\end{sdp}
\end{boxedminipage}
  \caption{SDP relaxation for the symmetric conductance of polymatroidal
    hypergraphs.}
\end{figure}

\begin{proof}
Let $G$ be the polymatroidal network obtained from $H$ using Lemma 
\ref{lem:H-to-G}. Fix an optimal solution to SDP \ref{sdp:u-h} on $G$. 
First note that for any $v \in V_H$, 
we have $\hrhoplus{\inG{v}}{\cdot}, \hrhominus{\inG{v}}{\cdot} = \infty$.  
Now, if $\dv{+}{\ell_2^2}{\vect{}}{\inG{v}} \neq 0$ or 
	$\dv{-}{\ell_2^2}{\vect{}}{\inG{v}} \neq 0$, then the cost of the SDP solution
would be $\infty$. Therefore, we have that  
\begin{equation} 
\label{eq:dv2-h}
\dv{+}{\ell_2^2}{\vect{}}{\inG{v}} = 0 \qquad \textrm{ and } \qquad 
\dv{-}{\ell_2^2}{\vect{}}{\inG{v}} = 0
\qquad \forall v \in V_H. 
\end{equation}
This implies that 
\begin{equation} 
\label{eq:dv3-h}
\sdpval = \frac{
\sum_{v \in \inG{V} \cup \inG{E}} \paren{\hrhoplus{v}{\dv{+}{\ell_2^2}{\vect{}}{v}} 
		+ \hrhominus{v}{\dv{-}{\ell_2^2}{\vect}{v}}}}
{\sum_{v \in \inG{E}} \deg_v \norm{\vect{v}}^2}
= \frac{
\sum_{v \in \inG{E}} \paren{\hrhoplus{v}{\dv{+}{\ell_2^2}{\vect{}}{v}} 
	+ \hrhominus{v}{\dv{-}{\ell_2^2}{\vect{}}{v}}}}
{\sum_{v \in \inG{E}} \deg_v \norm{\vect{v}}^2}.
\end{equation}
Since $x_i \defeq \inprod{\vect{i},g}$, we get from \ref{eq:dv2-h} that
\[ \dv{+}{\ell_2^2}{x}{\inG{v}} = 0 \qquad \textrm{ and } \qquad 
	\dv{-}{\ell_2^2}{x}{\inG{v}} = 0 \qquad \forall v \in V_H. \]
Let  $y$ be the vector obtained from $x$ using Lemma \ref{lem:shift}.
Since $y = x + c \allones$ for some $c \in \R$, we get that  
\[ \dv{+}{\ell_2^2}{y}{\inG{v}} = 0 \qquad \textrm{ and } \qquad 
	\dv{-}{\ell_2^2}{y}{\inG{v}} = 0 \qquad \forall v \in V_H. \]
Let $z$ be the vector obtained in the proof of Proposition \ref{prop:cheeger}. 
Since $z$ is either $y^+$ or $y^-$, we have that 
\begin{equation} 
\label{eq:dv1-h}
\dv{+}{\ell_1}{z^2}{\inG{v}} = 0 \qquad \textrm{ and } \qquad 
	\dv{-}{\ell_1}{z^2}{\inG{v}} = 0 \qquad \forall v \in V_H. 
\end{equation}
Therefore,
\begin{equation*}
\sum_{v \in \inG{V} \cup \inG{E}} \paren{\hrhoplus{v}{\dv{+}{\ell_1}{z^2}{v}} 
	+ \hrhominus{v}{\dv{-}{\ell_1}{z^2}{v}}}
= \sum_{v \in \inG{E}} \paren{\hrhoplus{v}{\dv{+}{\ell_1}{z^2}{v}} 
	+ \hrhominus{v}{\dv{-}{\ell_1}{z^2}{v}}}.
\end{equation*}
Following the proof of Proposition \ref{prop:cheeger}, we can show that
the set $S$ returned by the algorithm in this proposition satisfies
\begin{align}
\frac{\cut{G}{\delta^+(S)} + \cut{G}{\delta^-(S)}}{\sum_{v \in \inG{E} \cap S} \deg_v} 
& \leq \frac{\sum_{v \in \inG{E}} \paren{\hrhoplus{v}{\dvright{+}{\ell_1}{z^2}{v}} 
+ \hrhominus{v}{\dvleft{-}{\ell_1}{z^2}{v}}}}{\sum_{v \in \inG{E}} \deg_v z_v^2}
\nonumber \\ 
& \leq 32 \sqrt{
\frac{\sum_{v \in \inG{E}} \paren{\hrhoplus{v}{\dvright{+}{\ell_2^2}{x}{v}} 
	+ \hrhominus{v}{\dvleft{-}{\ell_2^2}{x}{v}}}}{\sum_{v \in \inG{E}} \deg_v x_v^2}}
	\nonumber \\
& = \bigo{\sqrt{\sdpval \log r}} \qquad \paren{\textrm{using \ref{eq:dv3-h}}} \nonumber \\
& = \bigo{\sqrt{\uexpnG{G} \log r}} \nonumber \\
& = \bigo{\sqrt{\uexpnG{H} \log r}} 
\qquad \paren{\textrm{using Corollary \ref{cor:H-to-G-expn}}} \label{eq:costT3}.
\end{align}
Using \ref{eq:dv1-h}, we also get that 
for any cut $(S,g)$ returned by the algorithm in Lemma 
\ref{lem:line-embedding}, $g^{-1}(\inG{v}) = \emptyset \ \forall v \in V_H$.

We now need to recover a set in the hypergraph $H$ from the set $S$ in $G$.
First, if there is some $\inG{e} \in S$ (resp. $V \setminus S$) such that 
all of its neighbours are in $V \setminus S$ (resp. $S$), then we move 
$\inG{e}$ to $V \setminus S$ (resp. $S$). Using the following fact, 
it follows that conductance will not increase by doing so.
\[ \frac{a - \epsilon}{b - \epsilon} \leq \frac{a}{b}
\qquad \forall a,b, \epsilon \textrm{ such that } 
0 < \epsilon \leq a \leq b \textrm{ and } \epsilon < b. \]
Henceforth, we will assume that if $\inG{e} \in S$ then it contains
at least on neighbour in $S$.
Let $T \subset V_H \cup E_H$ correspond to the set $S$, i.e., $T \defeq
\set{i \in V_H \cup E_H : \inG{i} \in S}$.

We get that 
\begin{align*}
\uexpn{H}{T} & =
\frac{\symcut{H}{T} + \symcut{H}{(V_H \cup E_H) \setminus T}}{\vol{H}{T}} \\ 
& = \frac{\cut{G}{\delta^+(S)} + \cut{G}{\delta^-(S)}}{\sum_{v \in \inG{E}\cap S} \deg_v} 
	& \textrm{(using Lemma \ref{lem:H-to-G-expn})} \\
& = \bigo{\sqrt{\uexpnG{H} \log r}} & \textrm{(using \ref{eq:costT3})}. 
\end{align*}
\end{proof}

\subsection{Hypergraph Conductance in \texorpdfstring{$r$}{r}-uniform hypergraphs}
\label{sec:hyper-exp}
Given a hypergraph $H = (V,E,w)$ with non-negative hyperedge weights $(w_e)_{e \in E}$, 
various notions of conductance have been studied; see Section \ref{sec:related}
for a brief survey.
We study a more natural notion of conductance of a set 
$T \subset V_H$ defined as follows.
\[ \psi_H(T) \defeq \frac{\sum_{e \in \delta_H(T)} w_e}
{\min \set{\sum_{\substack{e \in E: \\ e \cap T \neq \emptyset}} w_e,
\sum_{\substack{e \in E: \\ e \cap (V_H \setminus T) \neq \emptyset}} w_e }} \]
and $\psi_H \defeq \min_{T \subset V_H} \psi_H(T)$.
Here each hyperedge is counted at most once in the denominator of $\psi_H(T)$.

\begin{theorem}
\label{thm:alg2-h}
Let $H = \paren{V,E,w}$ be a hypergraph with $w : E \to \Rplus$ satisfying 
	$w_e \leq (1/4) \sum_{e' \in E} w_{e'} \ \forall e \in E$, and let
$r \defeq \max_{e \in E} \Abs{e}$.
There is a randomized polynomial time algorithm that outputs a set $T \subset V$ 
such that $\psi_H(T) = \bigo{\sqrt{\psi_H \log r}}$ with constant
probability.
\end{theorem}

\begin{proof}
For any $e \in E$, we define $\fplus{e}{A} = \fminus{e}{A} = w_e$ if $A \neq \emptyset$ and 
to be $0$ if $A= \emptyset$.
Then $\deg_e =\fplus{e}{e} + \fminus{e}{e} = 2 w_e \ \forall e \in E$.

For any subset $S \subset V_H \cup E_H$ satisfying 
$\forall e \in S, \ e \cap S \neq \emptyset$ and 
$\vol{H}{S} \leq \vol{H}{V}/2$.
Let $T_S \defeq S \cap V_H$ and define $S' \defeq S \cup \delta_H(T_S \cap V_H)$, 
i.e., add to $S$ all the hyperedges which are cut by the vertices in $T_S$.
Then
\begin{flalign}
& \sum_{e \in \delta_H(S' \cap V_H)} \paren{\fminus{e}{e \cap 
	\paren{V_H \setminus (V_H \cap S')}} + 
	\fplus{e}{e \cap \paren{V_H \setminus (V_H \cap S')}}} \nonumber \\
& = \sum_{e \in \delta_H(S \cap V_H)} \paren{\fminus{e}{e \cap 
	\paren{V_H \setminus (V_H \cap S)}} + 
	\fplus{e}{e \cap \paren{V_H \setminus (V_H \cap S)}}} \nonumber \\
& = 2 \sum_{e \in \delta_H(T_S)} w(e).
\label{eq:costT4}
\end{flalign} 

\begin{align}
\sum_{\substack{e \in E:\\ e \cap T_S \neq \emptyset}} 2 w(e) &
= \sum_{e \in S'} 2 w(e)
= \sum_{e \in S} \deg_e + \sum_{e \in S' \setminus S} 2 w(e) \nonumber \\
& \leq \sum_{e \in S} \deg_e + \paren{\sum_{e \in S} \deg_e}
\frac{\sum_{e \in \delta_H(S \cap V_H)} 2 w(e) }{\sum_{e \in S} \deg_e} \nonumber \\
& = \vol{H}{S} \paren{1 + \uexpn{H}{S}}.
\label{eq:costT5}
\end{align}

\begin{align}
\sum_{\substack{e \in E:\\ e \cap T_S \neq \emptyset}} 2 w(e) 
= \sum_{e \in S'} 2 w(e)
= \sum_{e \in S} \deg_e + \sum_{e \in S' \setminus S} 2 w(e) 
\geq \sum_{e \in S} \deg_e = \vol{H}{S} .
\label{eq:costT6}
\end{align}
Using \ref{eq:costT4}, \ref{eq:costT5} and \ref{eq:costT6}, we get
\begin{equation}
\label{eq:costT71}
\frac{1}{1 + \uexpn{H}{S}} \uexpn{H}{S} \leq \psi_H(T_S) \leq \uexpn{H}{S} . 
\end{equation}
Using \ref{eq:costT71} we get  
\begin{equation}
\label{eq:costT72}
\uexpnG{H} = \min_{S \subset V \cup E} \uexpn{H}{S} 
	\geq \min_{S \subset V \cup E} \psi_H(T_S)
	\geq \min_{T \subset V} \psi_H(T)
	= \psi_H.
\end{equation}
Since every $T \subseteq V$ is equal to $T_S$ for some $S \subseteq V \cup E$,
\begin{align}
\psi_H & = \min_{T \subset V} \psi_H(T) 
= \min_{T \subset V} \min_{\substack{S \subset V \cup E \\ T_S = T}} 
\psi_H(T_S) \nonumber 
\geq \min_{T \subset V} \min_{\substack{S \subset V \cup E \\ T_S = T}} 
\frac{1}{1 + \uexpn{H}{S}} \uexpn{H}{S}
& \textrm{(using \ref{eq:costT71})} \nonumber \\
& = \min_{S \subset V \cup E}
\frac{\uexpn{H}{S}}{1 + \uexpn{H}{S}} 
= \min_{S \subset V \cup E} \paren{1 - \frac{1}{1 + \uexpn{H}{S}}}
= 1 - \frac{1}{1 + \min_{S \subset V \cup E} \uexpn{H}{S}} \nonumber \\
& = 1 - \frac{1}{1 + \uexpnG{H}} = \frac{\uexpnG{H}}{1 + \uexpnG{H}} 
\geq \frac{1}{5} \uexpnG{H} & \paren{\textrm{using Claim \ref{claim:expn-eps}}}
	\label{eq:costT73}
\end{align}

Now, let $S$ be the set returned by the algorithm in Theorem \ref{thm:alg1-h}.
Then
\begin{align*}
\psi_H(T_S) & \leq \uexpn{H}{S} & \paren{\textrm{using \ref{eq:costT71}}} \\
	& = \bigo{\sqrt{\uexpnG{H} \log r}} &
	\paren{\textrm{using Theorem \ref{thm:alg1-h}}} \\
	& = \bigo{\sqrt{\psi_H \log r}} & \paren{\textrm{using \ref{eq:costT73}}} .
\end{align*}
\end{proof}

\begin{claim}
\label{claim:expn-eps}
There is a polynomial time algorithm to compute a set $A \subset V_H \cup E_H$
satisfying 
	\[ \uexpn{H}{A} \leq 4 .\]
\end{claim}
\begin{proof}
Recall our assumption that $w_e \leq (1/4) \sum_{e' \in E} w_{e'} \ \forall e \in E$.
Therefore, one can efficiently compute a set 
$A \subset V_H \cup E_H$ satisfying $\vol{H}{A} \in 
\left[\paren{\frac{1}{2} - \frac{1}{4}} \sum_{e\in E} w_e, 
\frac{1}{2} \sum_{e\in E} w_e \right]$.
Then 
\[ \uexpn{H}{A} \leq \frac{\vol{H}{E}}{\vol{H}{A}} 
	\leq \frac{\vol{H}{E}}{\paren{\frac{1}{4}} \vol{H}{E}} 
	\leq 4 .\]
\end{proof}

\subsection{Symmetric Vertex Expansion in Graphs}
\label{sec:vert-exp}
Let $G = (V,E,w)$ be a graph with vertex weights $w: V \to \Rplus$ satisfying
$w_v \leq (1/4) \sum_{v' \in V} w_{v'} \forall v \in V$.
For $S \subset V$, we define its inner boundary as the set of vertices in $S$ 
which have a neighbour outside $S$, i.e., $\nin{}{S} \defeq 
\set{v \in S : \exists u \in V \setminus S \textrm{ such that } \set{u,v} \in E}$,
and its outer boundary as the set of vertices in $V \setminus S$ which have a 
neighbour in $S$, i.e. $\nout{}{S} \defeq 
\set{v \in V \setminus S : \exists u \in S \textrm{ such that } \set{u,v} \in E}$.
The symmetric vertex expansion of a set $S \subset V$ is defined as
\[ \expn{}{G}{S} \defeq \frac{\sum_{v \in \nin{G}{S} \cup \nout{G}{S}} w_v}
	{\min \set{\sum_{v \in S} w_v, \sum_{v \in V \setminus S} w_v}} \]
and $\expnG{}{G} \defeq \min_{S \subset V} \expn{}{G}{S}$.
\cite{LouisRV13} gave a $\bigo{\sqrt{\opt \log r}}$ approximation bound
for symmetric vertex expansion where $r$ is the largest vertex degree.
This approximation guarantee can also be obtained as a corollary to 
Theorem \ref{thm:alg2-h}, Lemma \ref{lem:expn-psi} and Corollary \ref{cor:expn-psi}.

We will use the following reduction from \cite{LouisM16}.
\begin{definition}[\cite{LouisM16}]
Given a graph $G = (V,E,w)$ with $w: V \to \Rplus$, we construct a hypergraph
$H = (V,E',w')$ with $w' : E' \to \Rplus$ as follows.
For each $v \in V$, add a hyperedge $e_v \defeq \set{v} \cup N(v)$ to $E'$
and set $w'(e_v) \defeq w_v$.
\end{definition}
The proof of the following lemma closely follows the analogous proof in \cite{LouisM16}
relating $\expn{}{G}{S}$ and hypergraph conductance/expansion.
\begin{lemma}
\label{lem:expn-psi}
For any $S \subset V$, 
\[ \frac{1}{1 + \expn{}{G}{S}} \expn{}{G}{S} 
\leq \psi_H(S) \leq \expn{}{G}{S} . \]
\end{lemma}
\begin{proof}
Fix $S \subset V$. Consider $v \in \nin{G}{S}$ (resp. $v \in \nout{G}{S}$). 
Then $\exists u \in V \setminus S$ (resp. $\exists u \in S$ )
such that $\set{v,u} \in E$. Therefore $e_v \in \delta_H(S)$. 
Similarly, consider $e_v \in \delta_H(S)$ and $v \in S$ 
(resp. $v \in V \setminus S$). Then $e_v \cap (V \setminus S) \neq \emptyset$
(resp. $e_v \cap S \neq \emptyset$). Pick any $u \in e_v \cap (V \setminus S)$
(resp. $u \in e_v \cap S$). By construction of $H$, $\set{u,v} \in E$.
Therefore, $v \in \nin{G}{S}$ (resp. $v \in \nout{G}{S}$).
Therefore, 
\begin{equation}
\label{eq:costT7}
\sum_{v \in \nin{G}{S} \cup \nout{G}{S}} w_v = \sum_{e \in \delta_H(S)} 
w'(e) . 
\end{equation}

\begin{align}
\sum_{\substack{e \in E: \\ e \cap S \neq \emptyset}} w'(e)
& = \sum_{v \in S \cup \nout{G}{S}} w_v
= \sum_{v \in S} w_v + \sum_{v \in \nout{G}{S}} w_v \nonumber \\
& \leq \sum_{v \in S} w_v \paren{1 + \expn{}{G}{S}} 
\label{eq:costT8} 
\end{align}
and
\begin{equation}
\label{eq:costT9} 
\sum_{\substack{e \in E: \\ e \cap S \neq \emptyset}} w'(e)
= \sum_{v \in S \cup \nout{G}{S}} w_v \geq \sum_{v \in S} w_v 
\end{equation}
Using \ref{eq:costT7}, \ref{eq:costT8} and \ref{eq:costT9} we get  
\[ \frac{1}{1 + \expn{}{G}{S}} \expn{}{G}{S} 
\leq \psi_H(S) \leq \expn{}{G}{S} . \]

\end{proof}

\begin{corollary}
\label{cor:expn-psi}
\[ \frac{1}{5} \expnG{}{G} \leq \psi_H \leq \expnG{}{G} . \]
\end{corollary}

\begin{proof}
Lemma \ref{lem:expn-psi} directly implies $\psi_H \leq \expnG{}{G}$. To prove
the other direction, we can argue similar to Claim \ref{claim:expn-eps} to
assume w.l.o.g. that $\expn{}{G}{S} \leq 4$.
Then, using Lemma \ref{lem:expn-psi} finishes this proof.
\end{proof}

\section{Cheeger's Inequality for Directed Polymatroidal Networks}
\label{sec:cheeger}

We define $\gamma_2$, the spectral gap of a polymatroidal network, as follows.
\[ \gamma_2 \defeq \inf_{x : x \perp D \allones_V} \ral{G}{}{x} . \]
The spectral gap of undirected graphs (folklore) and hypergraphs \cite{ChanLTZ18}
are defined in the same way using the analogous definition of $\ral{}{}{x}$.

In the section we prove Theorem \ref{thm:cheeger}.
\tcheeger*

\begin{proof}
Let $\vect{} \perp D \allones_V$ be any vector with $\ral{}{}{\vect{}} \leq 1.1 \gamma_2$.
Using Proposition \ref{prop:cheeger} with $\vect{}$, we get that 
$\expnG{}{G} = \bigo{\sqrt{\ral{}{}{\vect{}}}} = \bigo{\sqrt{\gamma_2}}$.

To prove the other direction let $(S,g) = \argmin_{\substack{S,g \\ \vol{}{S} \leq \vol{}{V}/2}} \expn{}{G}{S,g}$.
Let us first consider the case where $\cut{}{\delta^+(S)} \leq \cut{}{\delta^-(S)}$.
Let $g^+ : \delta^+(S) \to V$ be the optimal assignment.
Define $x \in \R^{V \cup E}$ as follows. 
For $i \in V$, $x_i \defeq 1$ if $i\in S$ and other $0$ otherwise.
For $e = (u,v)$,
\[ x_e \defeq \begin{cases}
	x_u & \textrm{if } x_u = x_v \\
	x_v & \textrm{if } g(e) = u \\
	x_u & \textrm{if } g(e) = v \\
\end{cases} .\]
Then 
\begin{equation}
\label{eq:n3}
\expn{}{G}{S,g^+} = \frac{\sum_{v \in V} \paren{\hrhoplus{v}{\dvright{+}{\ell_2^2}{x}{v}} 
	+ \hrhominus{v}{\dvleft{-}{\ell_2^2}{x}{v}}}}{\sum_{v \in V} \deg_v x_v^2} 
	= \ral{}{}{x} .  
\end{equation}
However, $\inprod{x_V, D \allones_V}$ may not be equal to $0$ in general. 
Define $x' \defeq x - c \allones$ for an appropriate constant $c$ such that
$\inprod{x_V', D \allones_V} = 0$. 
Observe that 
\begin{equation}
\label{eq:n1}
\sum_{v \in V} \paren{\hrhoplus{v}{\dvright{+}{\ell_2^2}{x}{v}} 
	+ \hrhominus{v}{\dvleft{-}{\ell_2^2}{x}{v}}} =
	\sum_{v \in V} \paren{\hrhoplus{v}{\dvright{+}{\ell_2^2}{x'}{v}} 
	+ \hrhominus{v}{\dvleft{-}{\ell_2^2}{x'}{v}}} .
\end{equation}
Let us now calculate the value of $c$. 
\[ 0 = \inprod{x_V', D \allones_V} = \inprod{x_V - c \allones_V, D \allones_V} . \]
Therefore, using $\vol{G}{S} \leq \vol{G}{V}/2$, 
\[ c = \frac{\sum_{v \in V} \deg_v x_v}{\sum_{v \in V} \deg_v} 
	= \frac{\vol{G}{S}}{\vol{G}{V}} \in \left[0, \frac{1}{2} \right]. \]
\begin{align*} 
\sum_{v \in V} \deg_v (x_v')^2 & = (x_V')^T D x_V'
	= (x_V - c \allones_V)^T D (x_V - c \allones_V)
	= (x_V - c \allones_V)^T D x_V - c (x_V')^T D \allones_V \\
	& = x_V^T D x_V - c \allones_V^T D x_V + 0 \\
	& \geq \sum_{v \in V} \deg_v x_v^2 - \frac{1}{2} \paren{\sum_{v \in V} \deg_v x_v} = 
	\frac{1}{2} \sum_{v \in V} \deg_v x_v^2 
	\qquad \paren{\textrm{using } x_v \in \set{0,1} \ \forall v \in V} \\
\end{align*}
Using this and equation \ref{eq:n1}, we get 
\begin{equation} 
\label{eq:n2}
\ral{}{}{x} 
= \frac{\sum_{v \in V} \paren{ \hrhoplus{v}{\dvright{+}{\ell_2^2}{x}{v}} 
	+ \hrhominus{v}{\dvleft{-}{\ell_2^2}{x}{v}}}}{\sum_{v \in V} \deg_v x_v^2} 
\geq \frac{1}{2} \frac{\sum_{v \in V} \paren{\hrhoplus{v}{\dvright{+}{\ell_2^2}{x'}{v}} 
	+ \hrhominus{v}{\dvleft{-}{\ell_2^2}{x'}{v}}}}{\sum_{v \in V} \deg_v (x_v')^2} 
= \frac{1}{2} \ral{}{}{x'} . 
\end{equation}
Therefore, 
\begin{align*} 
\gamma_2 & = \inf_{y: y_V \perp D \allones_V} \ral{}{}{y} & \paren{\textrm{using the definition}} \\
	& \leq \ral{}{}{x'} & \paren{x_V' \perp D \allones_V \textrm{ by construction}} \\
	& \leq 2 \ral{}{}{x} & \paren{\textrm{using equation \ref{eq:n2}}} \\
	& \leq 2 \expnG{}{G} & \paren{\textrm{using equation \ref{eq:n3}}} .
\end{align*}

Next, let us consider the case where $\cut{}{\delta^+(S)} > \cut{}{\delta^-(S)}$.
Let $g^- : \delta^-(S) \to V$ be the optimal assignment.
Define $x \in \R^{V \cup E}$ as follows. 
For $i \in V$, $x_i \defeq - 1$ if $i\in S$ and other $0$ otherwise.
For $e = (u,v)$,
\[ x_e \defeq \begin{cases}
	x_u & \textrm{if } x_u = x_v \\
	x_v & \textrm{if } g(e) = u \\
	x_u & \textrm{if } g(e) = v \\
\end{cases} .\]
Note that $x \leq 0$.
\begin{align*}
\ral{}{}{x} & = \frac{\sum_{v \in V} \paren{\hrhoplus{v}{\dvright{+}{\ell_2^2}{x}{v}} 
	+ \hrhominus{v}{\dvleft{-}{\ell_2^2}{x}{v}}}}{\sum_{v \in V} \deg_v x_v^2} \\
& = \frac{\sum_{v \in V} \paren{\hrhoplus{v}{\dvleft{+}{\ell_2^2}{-x}{v}} 
	+ \hrhominus{v}{\dvright{-}{\ell_2^2}{-x}{v}}}}{\sum_{v \in V} \deg_v (-x_v)^2} 
 & \paren{\textrm{using Lemma \ref{lem:+-}}} \\
& = \expn{}{G}{S,g^-} .
\end{align*}
Let us now calculate the value of $c$. Using $\vol{G}{S} \leq \vol{G}{V}/2$, 
\[ c = \frac{\sum_{v \in V} \deg_v x_v}{\sum_{v \in V} \deg_v} 
	= \frac{-\vol{G}{S}}{\vol{G}{V}} \in \left[-\frac{1}{2},0 \right]. \]
Following the same steps as in the calculations as above, we get that 
$\gamma_2 \leq 2 \expnG{}{G}$ in this case as well. 
\end{proof}

\subsubsection*{Acknowledgements.}
We thank Yogeshwaran D. and Manjunath Krishnapur for providing us with 
a proof of Lemma \ref{lem:logd-approx}
and for allowing us to include it in this paper. 
CC thanks Kent Quanrud for helpful discussions.

\bibliographystyle{alpha}
\bibliography{submodular-cut}

\appendix

\section{Other Proofs}

\subsection{Line embeddings}
\label{sec:line-embedding}
We will use the following lemmas from \cite{ChekuriKRV15} (see Lemma 9, Lemma 10 and 
Remark 3 therein), we reproduce their proof here for the sake of completeness. 

\subsubsection{Proof of Lemma \ref{lem:line-embedding}}

\lineembedding* 

We will use the following lemma.
\begin{lemma}
\label{lem:line-embedding-step}
For any $v \in V$, and any monotone submodular functions $\rho^+ : \delta^+(v) \to \R$
	and $\rho^-: \delta^-(v) \to \R$.
\[ \hrhoplus{v}{\dvright{+}{\ell_1}{x}{v}} =
\int_0^{1} \rhoplus{v}{\set{e \in \delta^+(v):
	 x_v > t \geq x_e}} \dt \]
and
\[ \hrhominus{v}{\dvleft{-}{\ell_1}{x}{v}} = 
\int_0^{1} \rhominus{v}{\set{e \in \delta^-(v):
	 x_e > t \geq x_v}} \dt . \]
\end{lemma}

\begin{proof}
Fix any $v \in V$.
\begin{align*}
\hrhoplus{v}{\dvright{+}{\ell_1}{x}{v}} & = \int_0^1\rhoplus{v}{\set{e \in \delta^+(v):
	\dvright{+}{\ell_1}{x}{v}(e) \geq r}} \dr & \textrm{(using \ref{eq:lovext})} \\ 
& = \int_0^{1} \rhoplus{v}{\set{e \in \delta^+(v):
	\max\set{0, x_v - x_e} \geq r}} \dr & \paren{\textrm{using the definition of 
	$\dvright{+}{\ell_1}{x}{v}$}} \\
& = \int_0^{x_v} \rhoplus{v}{\set{e \in \delta^+(v):
	 x_v - x_e \geq r}} \dr & \paren{\textrm{since } r \geq 0 \textrm{ and } 
	 	x_v - x_e \leq x_v } \\
& = \int_0^{x_v} \rhoplus{v}{\set{e \in \delta^+(v):
	 x_v - r \geq x_e}} \dr \\
& = \int_0^{x_v} \rhoplus{v}{\set{e \in \delta^+(v):
	 t \geq x_e}} \dt & \paren{\textrm{using } t = x_v - r} \\
& = \int_0^{1} \rhoplus{v}{\set{e \in \delta^+(v):
	 x_v > t \geq x_e}} \dt .
\end{align*}
Similarly, we get 
\begin{align*}
\hrhominus{v}{\dvleft{-}{\ell_1}{x}{v}} & = \int_0^1\rhominus{v}{\set{e \in \delta^-(v):
	\dvleft{-}{\ell_1}{x}{v}(e) \geq r}} \dr & \textrm{(using \ref{eq:lovext})} \\ 
& = \int_0^{1} \rhominus{v}{\set{e \in \delta^-(v):
	\max\set{0, x_e - x_v} \geq r}} \dr & \paren{\textrm{using the definition of 
	$\dvright{+}{\ell_1}{x}{v}$}} \\
& = \int_0^{x_e} \rhominus{v}{\set{e \in \delta^-(v):
	 x_e - x_v \geq r}} \dr & \paren{\textrm{since } r \geq 0 \textrm{ and } 
	 	x_e - x_v \leq x_e }\\
& = \int_0^{x_e} \rhominus{v}{\set{e \in \delta^-(v):
	 x_e - r \geq x_v}} \dr \\
& = \int_0^{x_e} \rhominus{v}{\set{e \in \delta^-(v):
	 t \geq x_v}} \dt & \paren{\textrm{using } t = x_e - r} \\
& = \int_0^{1} \rhominus{v}{\set{e \in \delta^-(v):
	 x_e > t \geq x_v}} \dt . 
\end{align*}
\end{proof}

\begin{proof}[Proof of Lemma \ref{lem:line-embedding}]
If for any edge $(i,j) \in E$, $x_{(i,j)} > \max \set{x_i, x_j}$,
then setting $x_{(i,j)} = \max \set{x_i, x_j}$ will not increase the 
cost of the RHS. Similarly, if $x_{(i,j)} < \min \set{x_i, x_j}$, 
then setting $x_{(i,j)} = \min \set{x_i, x_j}$, will not increase the 
cost of the RHS. Therefore, we will henceforth assume that 
$\min \set{x_i, x_j} \leq x_{(i,j)} \leq \max \set{x_i, x_j}$ for 
all edges $(i,j) \in E$.

Using Lemma \ref{lem:line-embedding-step}, we get
\begin{flalign}
& \frac{\sum_{v \in V} \paren{\hrhoplus{v}{\dvright{+}{\ell_1}{x}{v}} 
	+ \hrhominus{v}{\dvleft{-}{\ell_1}{x}{v}}}}{\sum_{v \in V} w_v x_v} & \nonumber \\
& = \frac{\int_0^1 \sum_{v \in V} \paren{\rhoplus{v}{\set{e \in \delta^+(v):
	 x_v > t \geq x_e}} + \rhominus{v}{\set{e \in \delta^-(v):
	 x_v \leq t < x_e}}} \dt}{\int_0^1 \paren{\sum_{v \in V} w_v 
	\Ind{x_v > t}} \dt } & \nonumber \\
& \geq \min_{t \in [0,1]} \frac{\sum_{v \in V} \paren{\rhoplus{v}{\set{e \in \delta^+(v):
	 x_v > t \geq x_e}} + \rhominus{v}{\set{e \in \delta^-(v):
	 x_v \leq t < x_e}}}}{\paren{\sum_{v \in V} w_v 
	\Ind{x_v > t}}} . \label{eq:l1}
\end{flalign}
Note that the $\min$ above is well defined since there are at most $m + n$ 
different sets $S_t \defeq \set{a \in V \cup E: x_a > t}$ for $t \in [0,1)$.
Let $t^*$ be an optimal value of $t$ above, and define 
$g^+ : \delta^+(S_{t^*}) \to V$ as follows. For an edge $e = (u,v)$
where $u \in S_{t^*}$ and $v \notin S_{t^*}$,
\begin{equation}
\label{def:g+}
 g^+(e) \defeq \begin{cases} 
 u & \textrm{if } t^* \geq x_e \\
 v & \textrm{if } x_e > t^* \\
\end{cases} . 
\end{equation}
Then, 
\begin{align*}	
\frac{\cut{}{\delta^+(S_{t^*}),g^+}}{\wt{}{S_{t^*}}} & = \frac{\sum_{v \in V} 
	\paren{\rhoplus{v}{\delta^+(v) \cap (g^+)^{-1}(v)} 
	+ \rhominus{v}{\delta^-(v) \cap (g^+)^{-1}(v)}}}{\sum_{v \in S_{t^*}} w_v} \\
& = \frac{\sum_{v \in V} \paren{\rhoplus{v}{\set{e \in \delta^+(v):
	 x_v > t^* \geq x_e}} + \rhominus{v}{\set{e \in \delta^-(v):
	 x_v \leq t^* < x_e}}}}{\paren{\sum_{v \in V} w_v 
	\Ind{x_v > t^*}}} \\
& \leq \frac{\sum_{v \in V} \paren{\hrhoplus{v}{\dvright{+}{\ell_1}{x}{v}} 
	+ \hrhominus{v}{\dvleft{-}{\ell_1}{x}{v}}}}{\sum_{v \in V} w_v x_v} 
	\qquad \paren{\textrm{using \ref{eq:l1}}}.
\end{align*}
Suppose some $(u,v) \in E$ is cut by $S_{t^*}$ and we have $x_u = x_{(u,v)}$.
Since $u \in S_{t^*}$, we have $x_u > t^*$. Therefore, $x_{(u,v)} > t^*$.
From Definition \ref{def:g+}, we get that $g^+((u,v)) = v$.

The second part is proved similarly. Using Lemma \ref{lem:line-embedding-step}, we get
\begin{flalign}
& \frac{\sum_{v \in V} \paren{\hrhoplus{v}{\dvleft{+}{\ell_1}{x}{v}} 
	+ \hrhominus{v}{\dvright{-}{\ell_1}{x}{v}}}}{\sum_{v \in V} w_v x_v} & \nonumber \\
& = \frac{\int_0^1 \sum_{v \in V} \paren{\rhoplus{v}{\set{e \in \delta^+(v):
	 x_v \leq t < x_e}} + \rhominus{v}{\set{e \in \delta^-(v):
	 x_v > t \geq x_e}}} \dt}{\int_0^1 \paren{\sum_{v \in V} w_v 
	\Ind{x_v > t}} \dt } & \nonumber \\
& \geq \min_{t \in [0,1]} \frac{\sum_{v \in V} \paren{\rhoplus{v}{\set{e \in \delta^+(v):
	 x_v \leq t < x_e}} + \rhominus{v}{\set{e \in \delta^-(v):
	 x_v > t \geq x_e}}}}{\paren{\sum_{v \in V} w_v 
	\Ind{x_v > t}}} . \label{eq:l2}
\end{flalign}
Note that the $\min$ above is well defined since there are at most $m + n$ 
different sets $S_t \defeq \set{a \in V \cup E: x_a > t}$ for $t \in [0,1)$.
Let $t^*$ be an optimal value of $t$ above, and define 
$g^- : \delta^-(S_{t^*}) \to V$ as follows. For an edge $e = (v,u)$
where $u \in S_{t^*}$ and $v \notin S_{t^*}$,
\begin{equation}
\label{def:g-}
 g^-(e) \defeq \begin{cases} 
 u & \textrm{if } t^* \geq x_e \\
 v & \textrm{if } x_e > t^* \\
\end{cases} . 
\end{equation}
Then, 
\begin{align*}	
\frac{\cut{}{\delta^-(S_{t^*}),g^-}}{\wt{}{S_{t^*}}} & = \frac{\sum_{v \in V} 
	\paren{\rhoplus{v}{\delta^+(v) \cap (g^-)^{-1}(v)} 
	+ \rhominus{v}{\delta^-(v) \cap (g^-)^{-1}(v)}}}{\sum_{v \in S_{t^*}} w_v} \\
& = \frac{\sum_{v \in V} \rhoplus{v}{\set{e \in \delta^+(v):
	 x_v \leq t^* < x_e}} + \rhominus{v}{\set{e \in \delta^-(v):
	 x_v > t^* \geq x_e}}}{\paren{\sum_{v \in V} w_v 
	\Ind{x_v > t^*}}} \\
& \leq \frac{\sum_{v \in V} \paren{\hrhoplus{v}{\dvleft{+}{\ell_1}{x}{v}} 
	+ \hrhominus{v}{\dvright{-}{\ell_1}{x}{v}}}}{\sum_{v \in V} w_v x_v} 
	\qquad \paren{\textrm{using \ref{eq:l2}}}.
\end{align*}

\end{proof}

\subsubsection{Proof of Lemma \ref{lem:line-embedding-u}}

\lineembeddingu*

\begin{proof}
If for any edge $(i,j) \in E$, $x_{(i,j)} > \max \set{x_i, x_j}$,
then setting $x_{(i,j)} = \max \set{x_i, x_j}$ will not increase the 
cost of the RHS. Similarly, if $x_{(i,j)} < \min \set{x_i, x_j}$, 
then setting $x_{(i,j)} = \min \set{x_i, x_j}$, will not increase the 
cost of the RHS. Therefore, we will henceforth assume that 
$\min \set{x_i, x_j} \leq x_{(i,j)} \leq \max \set{x_i, x_j}$ for 
all edges $(i,j) \in E$.

Now, fix any $v \in V$.
Note that since for any $a,b \in \R$ we have 
$\Abs{a - b} = \max \set{0, a - b} + \max \set{0, b - a}$, we get that 
$\dv{+}{\ell_1}{x}{v} = \dvright{+}{\ell_1}{x}{v} + \dvleft{+}{\ell_1}{x}{v}$.
Using Lemma \ref{lem:summ-u}, we get that 
\[ \hrhoplus{v}{\dv{+}{\ell_1}{x}{v}} \geq \frac{1}{2} 
	\paren{\hrhoplus{v}{\dvright{+}{\ell_1}{x}{v}}
	+ \hrhoplus{v}{\dvleft{+}{\ell_1}{x}{v}}} . \]
Using Lemma \ref{lem:line-embedding-step}, we get 
\begin{equation}
\label{eq:cut+}
\hrhoplus{v}{\dv{+}{\ell_1}{x}{v}} = \frac{1}{2} 
	\paren{\int_0^{1} \paren{\rhoplus{v}{\set{e \in \delta^+(v):
	 x_v > t \geq x_e}} + \rhoplus{v}{\set{e \in \delta^+(v):
	 x_e > t \geq x_v}}} \dt }.
\end{equation}
Similarly, we get 
\begin{equation}
\label{eq:cut-u}
\hrhominus{v}{\dv{-}{\ell_1}{x}{v}} = \frac{1}{2} 
	\paren{\int_0^{1} \paren{\rhominus{v}{\set{e \in \delta^-(v):
	 x_v > t \geq x_e}} + \rhominus{v}{\set{e \in \delta^-(v):
	 x_e > t \geq x_v}}} \dt }.
\end{equation}

\begin{flalign}
& \frac{\sum_{v \in V} \paren{\hrhoplus{v}{\dv{+}{\ell_1}{x}{v}} 
	+ \hrhominus{v}{\dv{-}{\ell_1}{x}{v}}}}{\sum_{v \in V} w_v x_v} & \nonumber \\
& = \left(\int_0^{1} \sum_{v \in V} \left(\rhoplus{v}{\set{e \in \delta^+(v):
	 x_v > t \geq x_e}} + \rhoplus{v}{\set{e \in \delta^+(v):
	 x_e > t \geq x_v}} + \right. \right. \nonumber \\
& \left. \left.  \rhominus{v}{\set{e \in \delta^-(v):
	 x_v > t \geq x_e}} + \rhominus{v}{\set{e \in \delta^-(v):
	 x_e > t \geq x_v}} \right) \dt \right)/\paren{2 \int_0^1 
	 \paren{\sum_{v \in V} w_v \Ind{x_v > t}}} \dt \nonumber \\
& \geq \min_{t \in [0, 1)} \left(\sum_{v \in V} \left(\rhoplus{v}{\set{e \in \delta^+(v):
	 x_v > t \geq x_e}} + \rhoplus{v}{\set{e \in \delta^+(v):
	 x_e > t \geq x_v}} + \right. \right. \nonumber \\
& \left. \left.  \rhominus{v}{\set{e \in \delta^-(v):
	 x_v > t \geq x_e}} + \rhominus{v}{\set{e \in \delta^-(v):
	 x_e > t \geq x_v}} \right)  \right)/ \paren{2 \sum_{v \in V} w_v 
	\Ind{x_v > t}} \label{eq:l1-u} 
\end{flalign}
Note that the $\min$ above is well defined since there are at most $m + n$ 
different sets $S_t \defeq \set{a \in V \cup E: x_a > t}$ for $t \in [0,1)$.
Let $t^*$ be an optimal value of $t$ above, and define 
$g^+ : \delta^+(S_{t^*}) \to V$ and $g^- : \delta^-(S_{t^*}) \to V$
as follows. For an edge $e = (u,v)$
where $u \in S_{t^*}$ and $v \notin S_{t^*}$,
\[ g^+(e) \defeq \begin{cases} 
 u & \textrm{if } t^* \geq x_e \\
 v & \textrm{if } x_e > t^* \\
\end{cases} . \]
 For an edge $e = (u,v)$
where $u \notin S_{t^*}$ and $v \in S_{t^*}$,
\[ g^-(e) \defeq \begin{cases} 
 v & \textrm{if } t^* \geq x_e \\
 u & \textrm{if } x_e > t^* \\
\end{cases} . \]
Then, 
\begin{flalign*}	
& \frac{\cut{}{\delta^+(S_{t^*}),g^+} + \cut{}{\delta^-(S_{t^*}),g^-}}
	{\wt{}{S_{t^*}}} & \\
& = \left(\sum_{v \in V} \left(\rhoplus{v}{\set{e \in \delta^+(v):
	 x_v > t^* \geq x_e}} + \rhoplus{v}{\set{e \in \delta^+(v):
	 x_e > t^* \geq x_v}} + \right. \right. \nonumber & \\
& \left. \left. \rhominus{v}{\set{e \in \delta^-(v):
	 x_v > t^* \geq x_e}} + \rhominus{v}{\set{e \in \delta^-(v):
	 x_e > t^* \geq x_v}} \right) \right)/\paren{\sum_{v \in V} w_v 
	\Ind{x_v > t^*}} \\
& \leq 2 \frac{\sum_{v \in V} \paren{\hrhoplus{v}{\dv{+}{\ell_1}{x}{v}} 
	+ \hrhominus{v}{\dv{-}{\ell_1}{x}{v}}}}{\sum_{v \in V} w_v x_v} 
	\qquad \paren{\textrm{using \ref{eq:l1-u}}}.
\end{flalign*}

\end{proof}

\begin{lemma}
\label{lem:summ-u}
Let $\hat{\rho} : \R^S \to \R$ be the \lovasz~ extension of a monotone submodular
function $\rho : \set{0, 1}^S \to \R$. 
Let $x, x_1, x_2 \in \Rgeq^S$ such that $x = x_1 + x_2$. Then
$\hat{\rho}(x_1) + \hat{\rho}(x_2) \leq 2 \hat{\rho}(x)$.
\end{lemma}
\begin{proof}
Since $x - x_1 = x_2 \geq 0$, we have $x \geq x_1$. Similarly, we get $x \geq x_2$.
Using Fact \ref{fact:monotone-sub}, we that $\hat{\rho}(x) \geq 
\hat{\rho}(x_1), \hat{\rho}(x_2)$. 
Adding these two inequalities finishes the proof of the lemma.
\end{proof}

\subsection{Proof of Claim \ref{claim:sdpleqspar}}
\label{sec:sdpleqspar}

\SDPleqspar*

\begin{proof}
Let $\paren{S, V\setminus S}$ be an optimal cut.
W.l.o.g., let us assume that $\cut{}{\delta^+(S)} \leq \cut{}{\delta^-(S)}$
(if it isn't, we can denote $V \setminus S$ by $S$),
 and let $g^+: \delta^+(S) \to V$ be an optimal assignment.
Let $\alpha = 1/\sqrt{\paren{w(S) w(V \setminus S)}}$
Consider the following SDP solution. For $i \in V$, define
$\vect{i} := \alpha$ if $i \in S$ and $0$ otherwise, 
and $\vectone := \alpha$. For $(i,j) \in E$, define 
\[ \vect{(i,j)} := \begin{cases} \vect{i} & \textrm{if } \vect{i} = \vect{j} \\
	0 & \textrm{if } i \in S,\ j \in V \setminus S,\ g^+\paren{(i,j)} = i \\
	\alpha & 
	\textrm{if } i \in S,\ j \in V \setminus S,\ g^+\paren{(i,j)} = j\\
	\alpha &
	\textrm{if } j \in S,\ i \in V \setminus S \\
\end{cases} .\]

By construction,
\[ \norm{\vect{i} - \vect{j}}^2 = \begin{cases} \alpha^2 & \textrm{if }
			i \in S,\ j \in V \setminus S \textrm{ or } 
			i \in V \setminus S,\ j \in S \\
			0 & \textrm{otherwise} \end{cases}. \]
Therefore,
\[ \sum_{i,j \in V} w_i w_j \norm{\vect{i} - \vect{j}}^2 
 =  \sum_{i \in S, j \in V \setminus S} w_i w_j \alpha^2 
 = w(s) w(V \setminus S) \frac{1}{w(s) w(V \setminus S)} = 1. \]
Therefore, constraint \ref{eq:normalization} is satisfied. 
Since this SDP solution is a scaled $\{0,1\}$ solution, it satisfies the $\ell_2^2$ 
triangle inequalities (\ref{eq:l22}). Therefore, this SDP solution is feasible.

By construction, for any edge $(i,j)$ we have that $\vect{(i,j)}$ is 
either equal to $\vect{i}$ or equal to $\vect{j}$.
For any $(i,j) \in E$ with both $i,j \in S$ or both $i,j \in V \setminus S$, we have
$\ddist{\vect{i}, \vect{(i,j)}} = \ddist{\vect{(i,j)},\vect{i}} = 0$.
Therefore, this SDP solution contributes $0$ for the edges inside $S$ and $V \setminus S$.

For any $(i,j) \in \delta^-(S)$ with $j \in S, \ i \in V \setminus S$, we have 
\[ 2 \ddist{\vect{i}, \vect{(i,j)}} = 
\norm{\vect{i} - \vect{(i,j)}}^2 - \norm{\vectone - \vect{i}}^2
	+ \norm{\vectone - \vect{(i,j)}}^2 =
	\paren{0 - \alpha}^2 - \paren{\alpha - 0}^2 + \paren{\alpha - \alpha}^2
	= 0  \]
and 
\[ 2 \ddist{\vect{(i,j)},\vect{j}} = 
\norm{\vect{j} - \vect{(i,j)}}^2 - \norm{\vectone - \vect{(i,j)}}^2 
	+ \norm{\vectone - \vect{j}}^2 =
	\paren{\alpha - \alpha}^2 - \paren{\alpha - \alpha}^2 + \paren{\alpha - \alpha}^2
	= 0  . \]
Therefore, this SDP solution contributes $0$ for the edges in $\delta^{-}(S)$.

For any $(i,j) \in \delta^+(S)$ with $i \in S, \ j \in V \setminus S$, we have 
\begin{align*}
\ddist{\vect{i}, \vect{(i,j)}} & = \frac{1}{2}\paren{
	\norm{\vect{i} - \vect{(i,j)}}^2 - \norm{\vectone - \vect{i}}^2
	+ \norm{\vectone - \vect{(i,j)}}^2} \\
& = \frac{1}{2}\paren{
	\paren{\alpha- \vect{(i,j)}}^2 - \paren{\alpha - \alpha}^2
	+ \paren{\alpha - \vect{(i,j)}}^2} 
 = \paren{\alpha - \vect{(i,j)}}^2 \\
& = \begin{cases} \alpha^2 & \textrm{if } g^+((i,j)) = j \\
			0 & \textrm{if } g^+((i,j)) = i 
	\end{cases}
\end{align*}
and
\begin{align*}
\ddist{\vect{(i,j)},\vect{j}} & = \frac{1}{2}\paren{
	\norm{\vect{j} - \vect{(i,j)}}^2 - \norm{\vectone - \vect{(i,j)}}^2
	+ \norm{\vectone - \vect{j}}^2} \\
& = \frac{1}{2}\paren{
	\paren{0 - \vect{(i,j)}}^2 - \paren{\alpha - \vect{(i,j)}}^2
	+ \paren{\alpha - 0}^2}
	 = \alpha \vect{(i,j)} \\
& = \begin{cases} \alpha^2 & \textrm{if } g^+((i,j)) = i \\
			0 & \textrm{if } g^+((i,j)) = 0
	\end{cases}
\end{align*}
Therefore,
\begin{align*}
\sdpval & \leq \sum_{v \in V_G} \paren{\hrhoplus{v}{\dv{+}{\ddistsymb}{\vect{}}{v}} 
	+ \hrhominus{v}{\dv{-}{\ddistsymb}{\vect{}}{v}}} \\
 & = \alpha^2 \sum_{v \in V_G} \paren{\rhoplus{v}{\delta^+(S) \cap g^{-1}(v)} 
	+ \rhominus{v}{\delta^+(S) \cap g^{-1}(v)}} & \paren{\textrm{Using Fact \ref{fact:ax}}} \\
 & = \frac{\cut{}{S}}{w(S) w(V \setminus S)} = \spar_G .
\end{align*}
\end{proof}

\subsection{Proofs of Lemma \ref{lem:H-to-G} and Lemma \ref{lem:H-to-G-expn}}
\label{sec:H-to-G-proofs}

\subsubsection{Proof of Lemma \ref{lem:H-to-G}}
\HtoG*

\begin{proof}
Fix $\alpha, t \in \Rplus$ and $S\subset V_H$ such that $\cut{H}{S} = \alpha$
and $w_H(S) = t$. 
Consider the set 
\begin{equation}
\label{eq:defT}
	T \defeq \set{\inG{v}: v \in S} \cup \set{\inG{e}: e \subseteq S} 
	\cup \set{\inG{e} : e \in \delta_H(S) \textrm{ and } 
	\fplus{e}{e \cap (V_H \setminus S)} \leq \fminus{e}{e \cap S} } .
\end{equation}
Recall that 
\[ \cut{G}{\delta^+(T)}  = \sum_{v \in V_G} 
	\paren{\rhoplus{v}{\delta^+(v) \cap g_{\delta^+(T)}^{-1}(v)} 
	+ \rhominus{v}{\delta^-(v) \cap g_{\delta^+(T)}^{-1}(v)}}  \]
where $g$ is the optimal assignment of edges in $\delta^+(T)$ to $V_G$. 
Since $\rhoplus{\inG{v}}{\cdot}, \rhominus{\inG{v}}{\cdot} = \infty$  $\forall v \in V_H$,
$g_{\delta^+(T)}^{-1}(\cdot)$ will never be equal to $\inG{v}$ for some $v \in V_H$. 
Now, fix an $e \in E_H$ such that $\inG{e} \in T$.
Then 
\begin{equation}
	\label{eq:einS}
\delta^-\paren{\inG{e}} \cap 
	g_{\delta^+(T)}^{-1}\paren{\inG{e}} = \emptyset \textrm{ and }
	\delta^+\paren{\inG{e}} \cap 
	g_{\delta^+(T)}^{-1}\paren{\inG{e}} = \set{\inG{e}} \times 
	\set{\inG{i} : i \in e \cap \paren{V_H \setminus S}} 
\end{equation}
and
\begin{equation}
\label{eq:einS-}
\delta^-\paren{\inG{e}} \cap 
	g_{\delta^-(T)}^{-1}\paren{\inG{e}} = \set{\inG{i} : i \in e \cap \paren{V_H \setminus S}} 
	\times \set{\inG{e}} 
	\textrm{ and } \delta^+\paren{\inG{e}} \cap 
	g_{\delta^-(T)}^{-1}\paren{\inG{e}}  = \emptyset .
\end{equation}
Similarly, fix an $e \in E_H$ such that $\inG{e} \in N(T)$.
Then 
\begin{equation}
	\label{eq:einNS}
\delta^-\paren{\inG{e}} \cap 
	g_{\delta^+(T)}^{-1}\paren{\inG{e}} = \set{\inG{i} : i \in e \cap S} 
	\times \set{\inG{e}} \textrm{ and }
	\delta^+\paren{\inG{e}} \cap 
	g_{\delta^+(T)}^{-1}\paren{\inG{e}} = \emptyset
\end{equation}
and
\begin{equation}
	\label{eq:einNS-}
\delta^-\paren{\inG{e}} \cap 
	g_{\delta^-(T)}^{-1}\paren{\inG{e}} =  \emptyset \textrm{ and }
	\delta^+\paren{\inG{e}} \cap 
	g_{\delta^-(T)}^{-1}\paren{\inG{e}} = \set{\inG{e}} \times 
	\set{\inG{i} : i \in e \cap S}.
\end{equation}
Therefore, 
\begin{flalign*}
& \cut{G}{\delta^+(T)} \\
& = \sum_{v \in V_G} 
	\paren{\rhoplus{v}{\delta^+(v) \cap g_{\delta^+(T)}^{-1}(v)} 
	+ \rhominus{v}{\delta^-(v) \cap g_{\delta^+(T)}^{-1}(v)}} \\ 
& =  \sum_{\substack{e \in \delta_H(S)}}
	\paren{\rhoplus{\inG{e}}{\delta^+\paren{\inG{e}} \cap g_{\delta^+(T)}^{-1}(\inG{e})} 
	+ \rhominus{\inG{e}}{\delta^-\paren{\inG{e}} \cap g_{\delta^+(T)}^{-1}(\inG{e})}} \\ 
& =  \sum_{\substack{e \in \delta_H(S) \\ \inG{e} \in T}}
	\rhoplus{\inG{e}}{\set{\inG{e}} \times 
	\set{\inG{i} : i \in e \cap \paren{V_H \setminus S}}}
+ \sum_{\substack{e \in \delta_H(S) \\ \inG{e} \in N(T)}}
	\rhominus{\inG{e}}{\set{\inG{i} : i \in e \cap S} \times \set{\inG{e}}} \\
& \qquad \textrm{(Using \ref{eq:einS}, \ref{eq:einNS})}\\ 
& = \sum_{\substack{e \in \delta_H(S) \\ \inG{e} \in T}}
	\fplus{e}{e \cap \paren{V_H \setminus S}} 
+ \sum_{\substack{e \in \delta_H(S) \\ \inG{e} \in N(T)}}
	\fminus{e}{e \cap S} \qquad \paren{\textrm{Using Definition \ref{def:reduction}}}\\
& = \sum_{e \in \delta_H(S)}
	\min \set{\fplus{e}{e \cap \paren{V_H \setminus S}},
	\fminus{e}{e \cap S}} \qquad \textrm{(Using \ref{eq:defT})} \\ 
& = \cut{H}{\delta(S)} = \alpha.
\end{flalign*}
Therefore, $\min \set{\cut{G}{\delta^+(T)}, \cut{G}{\delta^-(T)}} \leq \alpha$.
Using \ref{eq:defT},
\begin{align*}
w_G(T) & = w_G\paren{\set{\inG{v}: v \in S} \cup \set{\inG{e}: e \subseteq S} 
	\cup \set{\inG{e} : e \in \delta_H(S) \textrm{ and } 
	\fplus{e}{e \cap (V_H \setminus S)}\leq \fminus{e}{e \cap S}}} \\
	& = w_G\paren{\set{\inG{v}: v \in S}} \\
	& = w_H(S) = t , 
\end{align*}
and similarly $w_G(V_G \setminus T) = w_H(V_H \setminus S) = n-t$.

To prove the other direction, fix a $T \subseteq V_G$ such that 
$\min \set{\cut{G}{\delta^+(T)},\cut{G}{\delta^-(T)}} = \alpha$.
First, let us consider the case when $\cut{G}{\delta^+(T)} \leq \cut{G}{\delta^-(T)}$.
Let $S \defeq \set{i \in V_H : \inG{i} \in T}$.
\begin{align*}
\alpha & =\cut{G}{\delta^+(T)} = \sum_{v \in V_G} 
	\paren{\rhoplus{v}{\delta^+(v) \cap g_{\delta^+(T)}^{-1}(v)} 
	+ \rhominus{\inG{e}}{\delta^-(v) \cap g_{\delta^+(T)}^{-1}(v)}} \\ 
& =  \sum_{\substack{\inG{e} \in T \cup N(T))}}
	\paren{\rhoplus{\inG{e}}{\delta^+\paren{\inG{e}} \cap g_{\delta^+(T)}^{-1}(\inG{e})} 
	+ \rhominus{\inG{e}}{\delta^-\paren{\inG{e}} \cap g_{\delta^+(T)}^{-1}(\inG{e})}} \\ 
& =  \sum_{\inG{e} \in T}
	\rhoplus{\inG{e}}{\set{\inG{e}} \times \set{\inG{i} : i \in e \cap \paren{V_H \setminus S}}} 
+ \sum_{\inG{e} \in N(T)}
	\rhominus{\inG{e}}{\set{\inG{i} : i \in e \cap S} \times \set{\inG{e}}} \\
& \qquad \textrm{(Using \ref{eq:einS} and \ref{eq:einNS})}\\ 
& = \sum_{\substack{\inG{e} \in T}}
	\fplus{e}{ e \cap \paren{V_H \setminus S}} 
+ \sum_{\substack{\inG{e} \in N(T)}}
	\fminus{e}{ e \cap S} \qquad \paren{\textrm{Using Definition \ref{def:reduction}}} \\
& \geq \sum_{e \in \delta_H\paren{S}}
	\min \set{\fplus{e}{ e \cap \paren{V_H \setminus S}},
	\fminus{e}{ e \cap S}} \\
& \qquad \paren{\textrm{Since }
	\set{\inG{e} : e \in \delta_H\paren{S}} \subset T \cup N(T)} \\ 
& = \cut{H}{\delta\paren{S}}.
\end{align*}
Moreover, 
\[ w_H(S) = w_H\paren{\set{i \in V_H : \inG{i} \in T}} 
= w_G\paren{T} = t , \]
and similarly $w_H(V_H \setminus S) = w_G(V_G \setminus T) = n-t$.

Next, let us consider the case when $\cut{G}{\delta^+(T)} > \cut{G}{\delta^-(T)} = \alpha$.
Let $S \defeq \set{i \in V_H : \inG{i} \in V_G \setminus T}$.
\begin{align*}
\alpha & = \cut{G}{\delta^-(T)} = \sum_{v \in V_G} 
	\paren{\rhoplus{v}{\delta^+(v) \cap g_{\delta^-(T)}^{-1}(v)} 
	+ \rhominus{\inG{e}}{\delta^-(v) \cap g_{\delta^-(T)}^{-1}(v)}} \\ 
& =  \sum_{\substack{\inG{e} \in T \cup N(T))}}
	\paren{\rhoplus{\inG{e}}{\delta^+\paren{\inG{e}} \cap g_{\delta^-(T)}^{-1}(\inG{e})} 
	+ \rhominus{\inG{e}}{\delta^-\paren{\inG{e}} \cap g_{\delta^-(T)}^{-1}(\inG{e})}} \\ 
& = \sum_{\inG{e} \in N(T)}
	\rhoplus{\inG{e}}{\inG{e}} \times \set{\inG{i} : i \in e \cap \paren{V_H \setminus S}}  \\
& \qquad 	+ \sum_{\inG{e} \in T} 
	\rhominus{\inG{e}}{\set{\set{\inG{i} : i \in e \cap \paren{V_H \setminus \paren{V_H \setminus S}}} 
	\times \set{\inG{e}} }} 
\qquad \textrm{(using \ref{eq:einS-} and \ref{eq:einNS-})}\\ 
& = \sum_{\substack{\inG{e} \in N(T)}}
	\fplus{e}{ e \cap \paren{V_H \setminus S}} 
	+ \sum_{\substack{\inG{e} \in T}} \fminus{e}{ e \cap S} 
	\qquad \paren{\textrm{using Definition \ref{def:reduction}}} \\
& \geq \sum_{e \in \delta_H\paren{S}}
	\min \set{\fplus{e}{ e \cap \paren{V_H \setminus S}},
	\fminus{e}{ e \cap S}} \qquad \paren{\textrm{using }\delta_H\paren{S} = \delta_H\paren{V_H \setminus S}} \\
& \qquad \paren{\textrm{Since }
	\set{\inG{e} : e \in \delta_H\paren{S}} \subset T \cup N(T)} \\ 
& = \cut{H}{\delta\paren{S}}.
\end{align*}
Moreover,
\[ w_H(S) = w_H\paren{\set{i \in V_H : \inG{i} \in V_G \setminus T}} 
= w_G\paren{V_G \setminus T} = n-t , \]
and similarly $w_H(V_H \setminus S) = w_G(T) = t$.

\end{proof}

\subsubsection{Proof of Lemma \ref{lem:H-to-G-expn}}
\HtoGexpn*

\begin{proof}
\begin{flalign*}
& \symcut{H}{T} + \symcut{H}{(V_H \cup E_H) \setminus T} \\
& = \sum_{e \in T \cap \delta_H(T \cap V_H)}\paren{
	\fminus{e}{e \cap \paren{(V_H \cup E_H) \setminus T}} 
	+ \fplus{e}{e \cap \paren{(V_H \cup E_H) \setminus T}}} \\ 
& + \sum_{e \in \paren{(V_H \cup E_H) \setminus T} \cap \delta_H(((V_H \cup E_H) \setminus T)
	\cap V_H)}\paren{\fminus{e}{e \cap T} + 
	\fplus{e}{e \cap  T}} \\
& = \sum_{e \in S}\paren{\rhominus{\inG{e}}{
	\delta^-\paren{\inG{e}} \cap \delta^-(S)} + \rhoplus{\inG{e}}{
	\delta^+\paren{\inG{e}} \cap \delta^+(S)}} \\
& + \sum_{e \in \paren{V_G \setminus S}}\paren{\rhominus{\inG{e}}{
	\delta^-\paren{\inG{e}} \cap \delta^-(V_G \setminus S)} + \rhoplus{\inG{e}}{
	\delta^+\paren{\inG{e}} \cap \delta^+(V_G \setminus S)}} \qquad 
	\textrm{(using Definition \ref{def:reduction})} \\
& = \sum_{e \in S}\paren{\rhominus{\inG{e}}{
	\delta^-\paren{\inG{e}} \cap \delta^-(S)} + \rhoplus{\inG{e}}{
	\delta^+\paren{\inG{e}} \cap \delta^+(S)}} \\
& + \sum_{e \in \paren{V_G \setminus S}}\paren{\rhominus{\inG{e}}{
	\delta^-\paren{\inG{e}} \cap \delta^+(S)} + \rhoplus{\inG{e}}{
	\delta^+\paren{\inG{e}} \cap \delta^-(S)}} \\ 
& \qquad \qquad \paren{\textrm{using } \delta^-(V_G \setminus S) = \delta^+(S),\ 
	\delta^+(V_G \setminus S)= \delta^-(S)}\\
	& = \cut{G}{\delta^-(S)} + \cut{G}{\delta^+(S)}. 
\end{flalign*}

\[ \vol{H}{T} = \sum_{e \in T} \deg_e =  
	\sum_{\inG{e} \in \inG{E}\cap S} \deg_{\inG{e}} 
	= \vol{G}{S \cap \inG{E_H}}. \]

\end{proof}

\end{document}